\documentclass[twocolumn,superscriptaddress,amsmath,amssymb,aps,prb]{revtex4-1}

\usepackage{graphicx}
\usepackage{amssymb}
\usepackage{amsmath}
\usepackage{mathrsfs}

\usepackage{multirow}
\usepackage{array,tabularx}
\usepackage{xcolor}
\usepackage{comment}
\usepackage{dcolumn}
\usepackage{bm}

\usepackage[colorlinks,urlcolor=blue,citecolor=blue,linkcolor=blue]{hyperref}

\newcommand{\ve}[1]{{\mathbf #1}}
\newcommand{\vect}[1]{{\mathbf #1}}
\newcommand{\Frac}[2]{\displaystyle\frac{#1}{#2}}

\renewcommand{\k}{{\bf k}}

\newcommand{\q}{{\bf q}}
\newcommand{\Q}{{\bf Q}}
\newcommand{\0}{{\bf 0}}
\newcommand{\sch}{Schr{\"o}dinger }

\newcommand{\area}{\mathcal{A}}
\newcommand{\ef}{E_{\text{F}}}
\newcommand{\efzero}{E_{\text{F}0}}
\newcommand{\Ryx}{R_{\text{X}}}
\newcommand{\ex}{E_{\text{X}}}
\newcommand{\exd}{E_{\text{X}}^{(d)}}
\newcommand{\exdzero}{E_{\text{X}}^{(d=0)}}
\newcommand{\exQd}{E_{\text{X}\textbf{Q}}^{(d)}}
\newcommand{\exQdef}{E_{\text{X}\textbf{Q}}^{(d,E_{\text{F}})}}
\newcommand{\exQdefmin}{E_{\text{X}\textbf{Q}_{\text{min}}}^{(d,E_{\text{F}})}}
\newcommand{\exQzerodef}{E_{\text{X}\textbf{0}}^{(d,E_{\text{F}})}}
\newcommand{\wCQef}{\omega_{\text{C}\textbf{Q}}^{(E_{\text{F}})}}
\newcommand{\wCQzeroef}{\omega_{\text{C}\textbf{0}}^{(E_{\text{F}})}}
\newcommand{\ax}{a_{\text{X}}}
\newcommand{\kf}{k_{\text{F}}}
\newcommand{\fs}{\text{FS}}

\newcommand{\bra}[1]{\langle{#1}|}
\newcommand{\ket}[1]{|{#1}\rangle}

\usepackage{float}
\usepackage{upgreek}
\usepackage{makeidx}
\usepackage{enumerate}
\DeclareGraphicsExtensions{.png,.jpg,.pdf,.mps,.gif,.bmp,.eps}
\usepackage{physics}

\begin{document}


\title{Extremely imbalanced two-dimensional electron-hole-photon systems}

\author{A. Tiene}
\email{antonio.tiene@uam.es}
\affiliation{Departamento de F\'isica Te\'orica de la Materia
  Condensada \& Condensed Matter Physics Center (IFIMAC), Universidad
  Aut\'onoma de Madrid, Madrid 28049, Spain}

\author{J. Levinsen}
\affiliation{School of Physics and Astronomy, Monash University, Victoria 3800, Australia}
\affiliation{ARC Centre of Excellence in Future Low-Energy Electronics Technologies, Monash University, Victoria 3800, Australia}

\author{M. M.~Parish}
\affiliation{School of Physics and Astronomy, Monash University, Victoria 3800, Australia}
\affiliation{ARC Centre of Excellence in Future Low-Energy Electronics Technologies, Monash University, Victoria 3800, Australia}

\author{A. H. MacDonald}
\affiliation{Department of Physics, The University of Texas at Austin,
  Austin, TX 78712, USA}

\author{J. Keeling}
\affiliation{SUPA, School of Physics and Astronomy, University of St Andrews, St Andrews, KY16 9SS, United Kingdom}

\author{F. M. Marchetti}
\email{francesca.marchetti@uam.es}
\affiliation{Departamento de F\'isica Te\'orica de la Materia
  Condensada \& Condensed Matter Physics Center (IFIMAC), Universidad
  Aut\'onoma de Madrid, Madrid 28049, Spain}

\date{November 20, 2019}

\begin{abstract}
We investigate the phases of two-dimensional electron-hole systems strongly coupled to a microcavity photon field in the limit of extreme charge imbalance. Using variational wave functions, we examine the competition between different electron-hole paired states for the specific cases of semiconducting III-V single quantum wells, electron-hole bilayers, and transition metal dichalcogenide monolayers embedded in a planar microcavity.
We show how the Fermi sea of excess charges modifies both the electron-hole bound state (exciton) properties and the dielectric constant of the cavity active medium, which in turn affects the photon component of the many-body polariton ground state. 
On the one hand, long-range Coulomb interactions and Pauli blocking of the Fermi sea promote electron-hole pairing with finite center-of-mass momentum, corresponding to an excitonic roton minimum. On the other hand, the strong coupling to the ultra-low-mass cavity photon mode favors zero-momentum pairs. 
We discuss the prospect of observing different types of electron-hole pairing in the photon spectrum. 
\end{abstract}


\maketitle

\section{Introduction}
\label{sec:Introduction}
Recent technological progress has led to precise and efficient
manipulation of electronic and optical properties of semiconductor solid-state devices.
In particular, it is now possible to study the interplay between strong light-matter coupling and electronic doping, with the prospect of generating and controlling novel strongly correlated phases 
involving photons, electron-hole pairs and an electron gas.
Electron-hole systems with charge imbalance are expected to display exotic pairing phenomena such as electron-hole pairs (excitons) with finite center-of-mass (CoM) momentum~\cite{Fulde-Ferrell_PR1964,Larkin-Ovchinnikov_1964,Casalbuoni-Nardulli_RMP2004,Kinnunen_RepPP2018,Parish_EPL2011,Cotlet_arxiv2018} and phase separation in momentum-space between zero-momentum pairs and excess fermions~\cite{Sarma_JPCS1963,Pieri_Tanatar_PRB2010,Varley_Lee_PRB2016}.
The finite CoM paired state is equivalent to the Fulde-Ferrell-Larkin-Ovchinnikov (FFLO) phase~\cite{Fulde-Ferrell_PR1964,Larkin-Ovchinnikov_1964}, a spatially modulated paired phase first proposed in the context of spin-imbalanced superconductors. 
However, a conclusive experimental observation of the FFLO phase remains elusive. 
It is therefore of particular interest to understand how such a state in an electron-hole system might be probed and controlled with light. 

One particularly interesting class of materials giving access to this regime are transition metal dichalcogenide (TMDC) monolayers~\cite{Mak2016}. 
These structures are characterized by distinctive excitonic effects, ascribed to two-dimensional (2D) confinement and weak dielectric screening of the carrier Coulomb interactions in the 2D limit~\cite{Berkelbach_PRB2013,Cudazzo_PRL2016}. 
Coupling between excitons and electrically injected free charge carriers has been recently demonstrated (see, e.g., Ref.~\onlinecite{Chernikov_PRL2015}), together with the realisation of electron-hole bilayers with independently tunable carrier densities~\cite{Wang_Nature_2019}.
Further, the large exciton-binding energies and strong light-matter
coupling of these materials grant the possibility of accessing polaritonic (exciton-photon superposition) phenomena at room
temperature. Indeed, the strong light-matter coupling regime has been recently achieved by embedding a TMDC monolayer into an optical microcavity~\cite{Menon_NatPh_2014}, enabling the observation of valley-polarized exciton-polaritons at room-temperature~\cite{Chen_NaturePh_2017,Zheng_NaturePh_2017,Liu_PRL_2017}.
Such structures provide an ideal environment in which to investigate the interplay between strong light-matter coupling and electronic doping because of the possibility of externally tuning the electron density and light-matter coupling~\cite{Sidler_NP2016}.

Imbalanced electron-hole photon systems may also be realized using III-V and II-VI semiconducting single or coupled quantum wells. In particular, double quantum wells with independent electrical contacts, that allow one to independently tune the electron and hole densities in each layer, have been realized~\cite{Sivan_PRL1992,Croxall_PRL2008,Seamons_PRL2009}.
Here, the 2D electron and hole gases are separated by a barrier which is high enough to prevent recombination while thin enough to allow inter-layer exciton formation.
Such gated structures have not yet been embedded in a  microcavity, so have not yet been studied with strong light-matter coupling. 
However, a 2D electron gas (2DEG) in a single quantum well embedded into a planar microcavity has been realized experimentally.  Indeed, such a device can be produced either optically as shown for GaAs-based quantum well structures in Refs.~\onlinecite{Rapaport_PRB2001,Qarry_SST2003,Bajoni_PRB2006} or by using
a modulation doped CdTe~\cite{Brunhes_PRB1999} and GaAs~\cite{Gabbay_PRL2007,Smolka_Science2014} quantum well embedded in a planar cavity.
In these structures, at low 2DEG density, the negatively charged exciton-polariton, corresponding to a superposition of a trion (two electrons
and one hole) and a cavity photon,
emerges as
a dominant feature in the system spectrum. At larger densities, this physics is expected to evolve into that of the Fermi-edge exciton-polariton, as described in Refs.~\onlinecite{mahan2013:many,Pimenov_PRB2017} and references therein.
A connected problem is that of the Fermi-polaron polaritons~\cite{Efimkin_PRB2017}. Recent spectroscopic measurements in a gate-tunable monolayer MoSe$_2$ embedded into an open microcavity structure~\cite{Sidler_NP2016}
have shown strong signatures of both  trion and polaron resonances, where a mobile impurity, e.g., an optically generated hole, is dressed by particle-hole excitations across the 2DEG Fermi surface.

In this paper, we discuss pairing effects in strongly carrier density imbalanced electron-hole
2D structures strongly coupled to a microcavity photon field. In the absence of light, it was previously shown that a sufficiently high density of excess charge causes the exciton energy to develop a roton minimum at finite CoM momentum~\cite{Parish_EPL2011,Cotlet_arxiv2018} that is related to the FFLO~\cite{Fulde-Ferrell_PR1964,Larkin-Ovchinnikov_1964} phase first proposed for conventional superconductors.
Here, we study how strong coupling to light affects this excitonic FFLO roton minimum. While long-range Coulomb interactions and Pauli blocking promote the formation of a finite CoM momentum bound state, the strong coupling to low mass cavity photons tends to suppress such a phase. Conversely, the formation of an FFLO phase suppresses the coupling to light.
We study the competition between these processes by deriving the phase diagram of the equilibrium extremely imbalanced electron-hole-photon system, focusing solely on pairing phenomena.
We show that the exciton mode is affected not only by the presence of the majority species Fermi sea, but, at the same time, the excess charge modifies  
the dielectric constant of the active medium and, thus, it also affects the energy of the cavity photon mode.
Consequences of this predicted energy shift of the photon mode in the presence of a Fermi sea can be observed by comparing structures with different light-matter coupling, e.g.,by embedding a different number of quantum wells into the planar cavity and thus in effect changing the Rabi splitting.

The paper is organized as follows: In Sec.~\ref{sec:model} we introduce the electron-hole-photon system we consider, its Hamiltonian, and the renormalization of the cavity photon energy in the presence of an active medium (Sec.~\ref{sec:renor}), i.e., a single or double quantum well embedded into a planar microcavity. In Sec.~\ref{sec:varia}, we describe the variational approach we employ to describe the extremely imbalanced electron-hole-photon system. The paired (bound) and normal (unbound) polariton phases we consider are described in Secs.~\ref{sec:bstat} and~\ref{sec:norma}, respectively.
Results for the case of III-V structures are described in Sec.~\ref{sec:resul}, while the specific case of doped TMDC monolayers embedded into a planar cavity is discussed in Sec.~\ref{sec:tmdcv}. Conclusions and perspectives are gathered in Sec.~\ref{sec:conc}. Additional information related to this work can be found in the appendices.

\begin{figure}
\centering
\includegraphics[width=0.9\columnwidth]{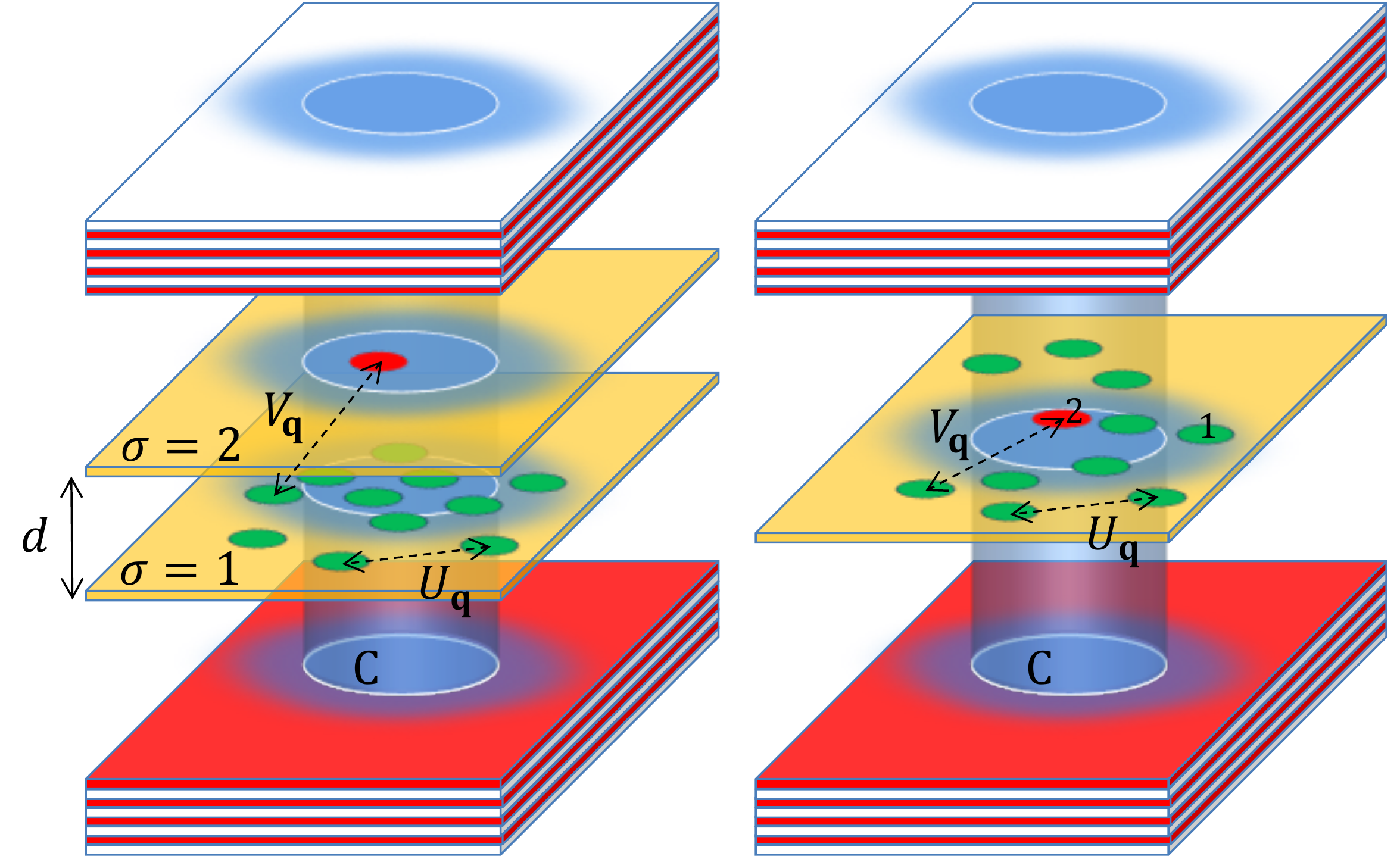}
\caption{Schematic representation of the 
system.
  Left panel: two quantum wells, labeled by the index
  $\sigma=1,2$ and separated by a distance $d$, form an electron-hole
  bilayer in the extremely imbalanced limit. The minority species
  belongs to the $\sigma=2$ layer, while the majority species at
  $\sigma=1$ forms an interacting Fermi sea --- the mass ratio $m_2/m_1$ establishes which layer is populated by either electrons or
  holes. $U_{\q}$ and $V_{\q}$ are respectively intra- and inter-species Coulomb interactions. The bilayer is located inside a
  planar cavity which confines the cavity photon mode
  ($\text{C}$). The (blue) shaded area represents the finite size
  external laser pump spot. Right panel: same set-up in a single
  quantum well geometry. Here, the majority $\sigma=1$ and minority
  $\sigma=2$ species belong to the same well.}
\label{fig:schem}
\end{figure}
%
\section{Model}
\label{sec:model}
We consider an electron-hole system in either a bilayer or a
single-layer geometry, embedded in a planar cavity. We consider the spin polarized case, where electrons and holes are in a single spin state (e.g., by introducing external magnetic field).  The system can be described by the
following Hamiltonian (in the following we set $\hbar=1$):
\begin{subequations}
\label{eq:hamge}
\begin{align}
\label{eq:hamil}
  \hat{H} &= \hat{H}_0 + \hat{H}_{\text{Coul}} +
  \hat{H}_{\text{e-h-C}}\\
\label{eq:freeh}
  \hat{H}_0 &=\sum_{\k\sigma} \left(\epsilon_{\k,\sigma} + \Frac{E_g}{2}\right)
  \hat{c}^\dag_{\k,\sigma} \hat{c}^{}_{\k,\sigma} + \sum_{\q}
  \nu_{\text{C}\q} \hat{a}_{\q}^{\dag} \hat{a}_{\q}^{} \\
\label{eq:Hcoulo}
  \hat{H}_{\text{Coul}} &= \sum_{\ve{k}\ve{k'}\ve{q},\sigma \sigma'}
    \frac{W_{\ve{q}}^{\sigma \sigma'}}{2\area} \hat{c}^\dag_{\ve{k},\sigma}
      \hat{c}^\dag_{\ve{k'},\sigma'} \hat{c}^{}_{\ve{k'}+\ve{q},\sigma'}
      \hat{c}^{}_{\ve{k}-\ve{q},\sigma}\\
  \hat{H}_{\text{e-h-C}} &= \frac{g}{\sqrt{\area}} \sum_{\k\q}
  \left(\hat{c}^\dag_{\frac{\q}2+\k,1} \hat{c}^\dag_{\frac{\q}2 - \k,2} \hat{a}_{\q}^{} +
    \text{h.c.}\right) \; ,
\label{eq:elhph}
\end{align}
\end{subequations}
where $\area$ is the system area. The Hamiltonian is defined
in terms of both matter
$\hat{c}^{}_{\k,\sigma=1,2}$ and cavity photon $\hat{a}_{\q}^{}$
operators.
The index $\sigma=1$ labels the majority species, and this has density $n_1$ and 
Fermi energy
\begin{equation}
 \ef=\Frac{\kf^2}{2m_1} = \Frac{2\pi}{m_1} n_1\; , 
\label{eq:fermi}
\end{equation}
where $k_F$ is the Fermi momentum. 
The index $\sigma=2$ indicates the minority species, for which we have at most one particle. Majority and minority particles correspond to distinct bands --- conduction band electrons and valence band holes.  In addition, for the bilayer geometry (left panel of Fig.~\ref{fig:schem}), the two species exist in two distinct quantum wells 
with transverse separation $d$.  For the single-layer geometry (right panel of Fig.~\ref{fig:schem}) both species live in the same quantum well.
The electrons and holes have dispersions
$\epsilon_{\k,\sigma} = \k^2/2m_{\sigma}$, where $m_\sigma$ is the mass, $\k$ is the two-dimensional (2D) momentum, and $E_g$ is the
band gap. For GaAs, the particle
mass ratio is $m_2/m_1 = 4$ in the case of a minority hole in a majority Fermi sea of electrons, or  $m_2/m_1 = 0.25$  for an electron in a Fermi sea of
holes.

The bare intra- and inter-species Coulomb interactions for a bilayer geometry involving two inorganic
quantum wells, such as III-V structures, are given respectively by
\begin{subequations}
\label{eq:coulo}
\begin{align}
\label{eq:coul1}
  W_{\ve{q}}^{\sigma \sigma} &= U_{\ve{q}} \equiv \frac{2\pi
    e^2}{\varepsilon q} \\
  W_{\ve{q}}^{12} &= W_{\ve{q}}^{21} = - V_\ve{q} \equiv - U_{\ve{q}} e^{-qd}\; ,
 \label{eq:coul2} 
\end{align}
\end{subequations}
where we use Gaussian units $4\pi \varepsilon_0=1$. In the absence of both doping and coupling to light, the Coulomb attraction between one electron and one hole leads to the Schr\"odinger equation for a 2D exciton~\cite{Yang_PRA1991}:
\begin{equation}
  \left(E 
  - \epsilon_{\k,1}-\epsilon_{\k,2}\right) \varphi_{\ve{k}} = -
  \sum_{\ve{k'}} \frac{V_{\ve{k}-\ve{k'}}}{\area}
  \varphi_{\ve{k'}} \; ,
\label{eq:eigex}
\end{equation}
where we measure the energy $E$ from the electron-hole band gap $E_g$.
Here, $\varphi_\k$ is the electron-hole wave function at relative momentum $\k$.
The negative energy solutions of this equation describe bound states and yield the exciton energies. Of particular interest is the 1$s$ exciton, with wave function $\Phi_{1s\k}^{(d)}$ and  binding energy $|\exd|$, where $E=\exd<0$ is the lowest energy eigenvalue of Eq.~\eqref{eq:eigex}.
For bilayers at a given separation $d$, the exciton properties can be found by numerically solving the \sch equation~\eqref{eq:eigex}. However, in the case of a single-layer geometry $d=0$, where $ V_\ve{q}=U_{\ve{q}}$, the \sch equation can be solved analytically, giving
the $1s$ exciton binding energy (or exciton Rydberg) 
in 2D in terms of the Bohr radius $\ax$ and the reduced mass $\mu=m_1 m_2/(m_1 + m_2)$:
\begin{align}
  \Ryx &=|\exdzero| = \Frac{e^2}{\varepsilon \ax}, & \ax &= \Frac{\varepsilon}{2\mu
    e^2} \; .
\label{eq:ebohr}
\end{align}
In this case, one
recovers the known
expression for the exciton wave function,
\begin{equation}
\Phi_{1s\k}^{(d=0)} = \Frac{\sqrt{8\pi} \ax}{[1 + (k\ax)^2]^{3/2}}\; .
  \label{eq:1swf}
\end{equation}

Electrons and holes couple with a strength $g$ to photons via the term $\hat{H}_{\text{e-h-C}}$~\eqref{eq:elhph}. The bare cavity photon dispersion is that of a passive cavity in the absence of the active medium (in our case a single or a double quantum well),
\begin{equation}
    \nu_{\text{C}\q} = \nu_{\text{C}\0} +
\Frac{\q^2}{2m_{\text{C}}}\; .
\end{equation}
We fix the photon mass $m_{\text{C}}$ to an experimentally relevant
value~\cite{Deng_RMP_2010}, $m_{\text{C}} \simeq 10^{-4} (m_1+m_2)$. 
In the presence of an active medium, the  cavity photon frequency is shifted by the coupling to matter excitations, as we discuss in the next section.

\subsection{Renormalization of the cavity photon energy}
\label{sec:renor}
In the presence of both light and matter degrees of freedom, the contact
coupling term of our model,
$\hat{H}_{\text{e-h-C}}$ in Eq.~\eqref{eq:elhph}, implies an ultraviolet
logarithmic divergence of the ground-state
energy~\cite{Levinsen_arxiv2019}. Since the
details of the high-momentum physics, such as the band curvature due to
the crystal lattice structure, are not included within our low-energy model, we will treat the
ultraviolet physics via the process of renormalization. This allows us
to deduce universal
properties of our system that are independent of 
microscopic details.

In order to see how the ultraviolet divergence emerges, it is instructive to
first consider the 
description of lower and upper polaritons within our model. To
this end, we follow Ref.~\onlinecite{Levinsen_arxiv2019} and consider the most
general superposition of a cavity photon at normal incidence and an
electron-hole pair:
\begin{equation}
  \ket{\Psi_{\0}} = \left(\sum_{\ve{k}}
    \Frac{\varphi_{\ve{k}\0}}{\sqrt{\area}}  \hat{c}^\dag_{\ve{k},1} \hat{c}^\dag_{ -
      \ve{k},2} +
    \alpha_{\ve{0}} \hat{a}_{\0}^{\dag}\right) \ket{0} \; .
\label{eq:polarQ0}
\end{equation}
Here, $\ket{0}$ is the vacuum state (i.e., a filled valence band, so a vacuum for valence band holes), $\alpha_\0$ is the photon
amplitude, and $\varphi_{\k\0}$ is the electron-hole wave function at
relative momentum $\k$ and zero CoM momentum.
Minimizing $\bra{\Psi_{\0}} (\hat{H} -E -E_g) \ket{\Psi_{\0}}$ with respect to the complex amplitudes
$\varphi_{\ve{k}\0}$ and $\alpha_{\0}$, we obtain the
coupled eigenvalue equations for the energy $E$ (measured with respect to the band gap $E_g$) 
of the polariton state:
\begin{subequations}
\label{eq:eige0}
\begin{align}
\label{eq:pola1}
  &\left(E  
  - \epsilon_{\k,1}-\epsilon_{\k,2}\right) \varphi_{\ve{k}\0} = -
  \sum_{\ve{k'}} \frac{V_{\ve{k}-\ve{k'}}}{\area}
  \varphi_{\ve{k'}\0} + g \alpha_{\ve{0}}
\\
    &\left(E 
  - \nu_{\text{C}\0} + E_g\right)
    \alpha_{\ve{0}} = \Frac{g}{\area} \sum_{\ve{k}}
    \varphi_{\ve{k}\0} \; .
\label{eq:pola2}
\end{align}
\end{subequations}
Inserting Eq.~\eqref{eq:pola1} in Eq.~\eqref{eq:pola2} and rearranging,
we obtain
\begin{multline}
  \left(E - \nu_{\text{C}\0} + E_g + \frac{g^2}{\area}\sum_\k\frac{1}{-E+ \epsilon_{\k,1}+\epsilon_{\k,2}}\right)\alpha_\0 \\
 =\frac{g}{\area^2}\sum_{\k,\k'} \frac{V_{\k-\k'}\varphi_{\k'\0}}{-E+\epsilon_{\k,1}+\epsilon_{\k,2}} \; .
\label{eq:renorm}
\end{multline}
The sum on the left hand side of this equation diverges.
If we introduce an ultraviolet momentum cutoff $\Lambda$, the divergence is logarithmic in $\Lambda$.
In contrast, the right hand
side is finite when $\Lambda\to\infty$~\cite{Levinsen_arxiv2019}. 
One can easily check this in the exciton limit --- i.e., where $g$ is small --- using 
the explicit form of the exciton wave function in a single quantum well, Eq.~\eqref{eq:1swf}. As a
consequence, the photon amplitude $\alpha_{\0}$ must approach zero as $1/\ln\Lambda$ for energies $E+E_g \sim\nu_{\rm{C}\0}$, which is a signature that the
photon frequency shifts in the presence of an active medium.
To have finite answers, it is therefore necessary that $\nu_{\text{C}\0}$ also diverges as $\ln \Lambda$ ---
i.e., we should write quantities in terms of the renormalized (finite and measurable) photon energy~\cite{Levinsen_arxiv2019} $\omega_{\text{C}\0}$ as follows:
\begin{equation}
  \omega_{\text{C}\0} = \nu_{\text{C}\0} 
  - \Frac{g^2}{\area} \sum_{\k} \Frac{1}{ -\exd +\epsilon_{\k,1}+\epsilon_{\k,2}} \; ,
\label{eq:renor}
\end{equation}
correct to logarithmic accuracy.
Here, we have taken the 1$s$ exciton binding energy to be the relevant energy scale $E \simeq \exd$, since we are considering the scenario where the photon is resonantly coupled to the 1$s$ exciton. 
One can thus define the renormalized photon-exciton detuning at zero momentum
\begin{equation}
  \delta =  \omega_{\text{C}\0} - (\exd + E_g)\; ,
\label{eq:detun}
\end{equation}
where $\exd+E_g$ is the actual exciton energy that would be measured spectroscopically.
Hence, the logarithmic divergence of~Eq.~\eqref{eq:renor} exactly
compensates the divergence appearing in Eq.~\eqref{eq:renorm} such that,
when this is expressed in terms of the dressed photon energy $\omega_{\text{C}\0}$ rather than
the bare photon energy $\nu_{\text{C}\0}$, one obtains convergent
cut-off independent results.

While the photon frequency is renormalized in the presence of an active
medium, the Rabi splitting between the lower and upper polaritons remains finite~\cite{Levinsen_arxiv2019}. In the limit $g \ll a_X \Ryx$ this splitting can be written as:
\begin{equation}
  \Omega = \Frac{2 g}{\area} \sum_{\k} \Phi_{1s\k}^{(d)} \; .
\label{eq:rabic}
\end{equation}
Here, $\Phi_{1s\k}^{(d)}$ is the ground state wave function of
Eq.~\eqref{eq:eigex} at layer separation $d$. Indeed, we see from
Eq.~\eqref{eq:renorm} that, had we chosen to compensate the logarithmic
divergence by taking $g\sim 1/\sqrt{\ln\Lambda}$, then the right hand
side of that equation would go to zero as $\Lambda\to\infty$, and we
would have had no coupling between light and matter.

One can show\cite{Levinsen_arxiv2019} that the implementation
of the renormalization scheme in Eqs.~\eqref{eq:pola1}
and~\eqref{eq:pola2} recovers the coupled exciton-photon oscillator
model in the limit $g \ll \ax \Ryx$. The generalization to finite
momentum is straightforward\cite{Levinsen_arxiv2019}, and one finds
that the lowest eigenvalue $E+E_g$ of Eqs.~\eqref{eq:pola1}
and~\eqref{eq:pola2} well
matches the one-particle lower polariton (LP) energy expression
coming from the two-level coupled oscillator model,
\begin{multline}
  \omega_{\text{LP}\Q} = \Frac{\omega_{\text{C}\Q} + 
  (\exQd + E_g)}{2} \\
    -
  \frac{1}{2} \sqrt{[\omega_{\text{C}\Q} - ( \exQd + E_g)]^2 +
    \Omega^2} \; ,
\label{eq:lpcou}    
\end{multline}
where the exciton is assumed to be a structureless particle.
Here, $\exQd = \exd + \Q^2/2(m_1+m_2)$ and
$\omega_{\text{C}\Q} = \omega_{\text{C}\0} + \Q^2/2m_{\text{C}}$.
As shown in Ref.~\onlinecite{Levinsen_arxiv2019}, the definitions
of the effective detuning, Eq.~\eqref{eq:renor}, and Rabi
splitting, Eq.~\eqref{eq:rabic}, represent a first order approximation in the expansion parameter $g \ll \ax \Ryx$ to the experimentally measured
detuning and Rabi splitting. An effort to obtain a better estimate of both parameters and a comparison with the approximation carried out here is discussed in  App.~\ref{app:fitti}. There, we employ a definition of detuning and Rabi splitting which is similar to a possible experimental procedure. In particular, we obtain their values by least squares fitting to match the LP dispersion obtained from a coupled oscillator model, Eq.~\eqref{eq:lpcou}. In this way, we find that the differences between the fitted parameters and those defined in Eqs.~\eqref{eq:detun} and~\eqref{eq:rabic} are small. This implies only small quantitative changes in our results below when we push our results beyond the $g\ll \Ryx \ax$ validity regime of Eqs.~\eqref{eq:detun} and~\eqref{eq:rabic}.

As in
Ref.~\onlinecite{Levinsen_arxiv2019}, the renormalization procedure we consider is defined for the case  at zero gating/doping ($\ef=0$). We  then increase the density of
majority particles while keeping the other parameters fixed. Interestingly, 
a TMDC monolayer flake embedded in a planar cavity offers the
possibility to measure independently the renormalized photon energy
$\omega_{\text{C}\0}$ and compare it to the bare value
$\nu_{\text{C}\0}$. In this structure, the TMDC flake has a
reduced size compared to the planar cavity, and thus there are regions
where the cavity mode is passive and does not couple to the active
medium~\cite{Menon_NatPh_2014}. Such a measurement would reveal that in the real system, the energy correction due to dressing is actually finite.  That is, an effective UV cutoff does indeed exist associated with the nature of electronic states at large momenta; however this cutoff is a high energy effect, beyond the scope of our low-energy Hamiltonian.

The definitions we adopt above for renormalization --- i.e., how we choose to calibrate the definitions of detuning --- match what we anticipate as a typical experimental protocol. Specifically, it corresponds to a process where, in the absence
of gating/doping, i.e., at $\ef=0$, one deduces the photon-exciton detuning
$\delta$ and the Rabi splitting $\Omega$ by fitting the
single-particle polariton dispersion measured in the optical pumping
linear regime via a coupled oscillator model. After fixing these experimental
conditions, one then increases $\ef$ by doping or gating. 
 The Rabi splitting $\Omega$ can be changed by
considering 
microcavities with different numbers of embedded quantum
wells~\cite{Brodbeck_PRL2017}.
The detuning
$\delta$ can be changed because of the cavity mirror wedge and thus by
changing the location of the optical pump spot. Crucially, the value of $\delta$ we use is defined as that measured in the absence of doping or gating before increasing $\ef$ --- i.e., we assume a definition of $\delta$ that does not vary with doping.

\subsection{Screening} 
\label{sec:scree}
In writing the Coulomb interaction above, we so far considered the bare Coulomb interaction.  However, as we consider a system with electronic doping, these electrons can screen and thus modify the Coulomb interaction.
As explained in Sec~\ref{sec:varia}, the screening of Coulomb interactions causes, in the absence of photons, 
a transition from bound to unbound excitonic states when
the majority species density
increases~\cite{Parish_EPL2011}. With the aim of including the possibility of describing the binding-unbinding transition, 
we will present 
results for both the unscreened case, and for screened Coulomb interactions within the static random phase approximation (RPA). In RPA, the intraspecies potential reads
\begin{subequations}
\begin{align}
  U^{sc}_{\ve{q}} &= \frac{U_\ve{q}} {1 - U_\ve{q}\Pi_{1}(\ve{q})}\\
  \Pi_{1}(\ve{q}) & = \frac{N_s m_1}{2\pi} \left[\frac{\sqrt{q^2 - 4
        \kf^2}}{q} \theta (q-2\kf)-1\right] \; ,
\end{align}
\end{subequations}
with $N_s = 1$ for the spin polarized case. As before, the
interspecies potential is then found by, $V^{sc}_{\ve{q}} = U^{sc}_{\ve{q}}
e^{-qd}$. We expect RPA to provide a good approximation when the exciton Bohr radius greatly exceeds the interparticle spacing of the majority species, i.e., $\ax^2 n_1 \gg 1$.  In the opposite limit, $\ax^2 n_1 \ll 1$, screening is negligible.
With this in mind, unscreened and RPA screened interactions represent extreme limiting cases, thus allowing us to place a bound on the effect of screening in a realistic material.

\section{Variational approach to the imbalanced electron-hole-photon system}
\label{sec:varia}
As described in the introduction, the aim of this paper is to understand how strong light-matter coupling affects the transition from excitons with zero to finite CoM momentum, as one varies the majority species density. To address this, we focus on the extreme limit, where there is a  single minority particle $\sigma=2$ interacting with a Fermi liquid of majority particles $\sigma=1$
via both  Coulomb attraction and the cavity mode.
To determine the zero temperature phase diagram,  we find the ground state by a variational approach. 
The variational state we consider describes a superposition of a photon and an electron-hole pair, on top of a Fermi sea of majority particles,
$\ket{\text{FS}} = \ket{\text{FS}}_1 \otimes \ket{0}_2 \otimes
\ket{0}_{\text{C}}$:
\begin{equation}
  \ket{\Psi_{\ve{Q}}} = \left(\sum_{\ve{k}>\kf}
    \Frac{\varphi_{\ve{k}\ve{Q}}}{\sqrt{\area}} \hat{c}^\dag_{\ve{k},1} \hat{c}^\dag_{\ve{Q} -
      \ve{k},2}  +
    \alpha_{\ve{Q}} \hat{a}_{\ve{Q}}^{\dag}\right) \ket{\text{FS}} \; .
\label{eq:polar}
\end{equation}
Here, $\varphi_{\ve{k}\ve{Q}}$ and $\alpha_{\ve{Q}}$ are the excitonic
and photonic variational parameters, respectively, and 
the normalisation condition requires that $\langle \Psi_{\ve{Q}}
\ket{\Psi_{\ve{Q}}} = \area^{-1}\sum_{\ve{k}>\kf}
|\varphi_{\ve{k}\ve{Q}}|^2 + |\alpha_{\ve{Q}}|^2 = 1$. 
The momentum $\Q$ is the CoM momentum of the polaritonic bound state, while the label $\vect{k}$ denotes the relative electron-hole  momentum.
Pauli blocking forbids occupation of all majority particle states below the Fermi momentum $\kf$, and we use the notation $\sum_{\ve{k}>\kf}$ to indicate summation  over allowed states.

\subsection{FF and SF bound states} 
\label{sec:bstat}
In the following, we will refer to the many-body polaritonic bound
state with finite CoM momentum $\ket{\Psi_{\Q\ne\0}}$ as the
FF state. Note that we use  the notation FF rather than FFLO because the pairing wave-function we consider is a single plane-wave, and thus it does not have any spatial modulation of density~\cite{Fulde-Ferrell_PR1964}. If we would consider increasing the density of minority particles, we expect a smooth evolution from the finite $\Q$
bound state we describe here to a modulated coherent FFLO 
paired phase~\cite{Varley_Lee_PRB2016}.
In the absence of cavity photons, the finite $\Q$ bound state for a single impurity has already been analysed for
GaAs~\cite{Parish_EPL2011} and TMDC~\cite{Cotlet_arxiv2018}
structures, where it was predicted to occupy a sizeable region of the
phase diagram. For an imbalanced state of electron-hole
bilayers, with a non-vanishing density of minority particles, a 
FFLO phase was also described in
Refs.~\onlinecite{Pieri_Tanatar_PRB2010,Varley_Lee_PRB2016}.

Also by analogy to the terminology used to describe the states at non-zero minority density, we refer to the zero CoM
momentum bound state $\ket{\Psi_{\0}}$ as the superfluid (SF) state. For a finite minority particle density, the SF state is an excitonic condensate where pairing occurs for a balanced fraction of electrons and holes at zero CoM momentum (but finite relative momentum), while the excess majority species occupies a Fermi sea around $\k=\0$.

To find which state occurs in the presence of coupling to photons,
we minimize $\bra{\Psi_{\ve{Q}}} (\hat{H} - E_{\Q} -E_g)
\ket{\Psi_{\ve{Q}}}$ with respect to the complex amplitudes
$\varphi_{\ve{k}\ve{Q}}$ and $\alpha_{\ve{Q}}$~\eqref{eq:polar}. This gives the
coupled eigenvalue equations
\begin{subequations}
\label{eq:eigen}
\begin{align}
\label{eq:eige1}
  &\left(E_{\Q}  - \xi_{\k\Q}\right) \varphi_{\ve{k}\ve{Q}} = -
  \sum_{\ve{k'}>\kf} \frac{V_{\ve{k}-\ve{k'}}}{\area}
  \varphi_{\ve{k'}\ve{Q}} + g \alpha_{\ve{Q}}\\
    &\left(E_{\Q}  - \nu_{\text{C}\vect{Q}} + E_g\right)
    \alpha_{\ve{Q}} = \Frac{g}{\area} \sum_{\ve{k}>\kf}
    \varphi_{\ve{k}\ve{Q}} \; .
\label{eq:eige2}
\end{align}
\end{subequations}
The lowest energy eigenvalue $E_{\Q}$  represents the binding energy of a bound lower polariton state in the presence of a Fermi sea, accounting for the modification of the exciton wavefunction both by light-matter coupling and by Pauli blocking. 
Here, $\xi_{\k\Q} = \epsilon_{\ve{Q}-\ve{k},2} + \epsilon_{\ve{k},1} -
\frac{1}{\area} \sum_{\ve{k}'<\kf} U_{\ve{k}-\ve{k'}}$ includes the exchange correction to the electron dispersion.
Note again that we define the energy $E_{\Q}$ with
respect to the band gap energy $E_g$; further we neglect the energy of the interacting Fermi sea $\ket{\fs}$,
$\mathscr{E}_{\text{FS}} = \sum_{\k < \kf} [\epsilon_{\k,1} + E_g/2 - \sum_{\k' < \kf} U_{\k-\k'}/(2\area)]$, because we are interested in comparing $E_{\Q}$ with that of the normal state, which also includes $\mathscr{E}_{\text{FS}}$ (see Sec.~\ref{sec:norma}).

In the absence of photons, we set $g=0$ in Eq.~\eqref{eq:eige1} and obtain the energy $E_{\Q}=\exQdef$ of a many-body exciton state in the presence of a Fermi sea as the lowest energy solution of the \sch equation
\begin{equation}
  \left(E_{\Q}  - \xi_{\k\Q}\right) \varphi_{\ve{k}\ve{Q}} = -
  \sum_{\ve{k'}>\kf} \frac{V_{\ve{k}-\ve{k'}}}{\area}
  \varphi_{\ve{k'}\ve{Q}} \; .
\label{eq:eige3}
\end{equation}
At zero doping and for a single layer, $E_{\text{X}\0}^{(d=0,\ef=0)}=\exdzero=-\Ryx<0$ as given in Eq.~\eqref{eq:ebohr}.
The eigenvalue problem in Eq.~\eqref{eq:eige3} has been solved numerically for GaAs electron-hole structures in Ref.~\onlinecite{Parish_EPL2011}. There, it was found that, when increasing the majority particle density, the many-body excitonic state eventually acquires a finite CoM momentum $\Q$, as this state reduces the kinetic energy cost. As such, this FF-like state induced by Pauli blocking is  favored when the minority particle is lighter. Further, as also discussed below, the long-range nature of the Coulomb interaction also stabilize the finite $\Q$ exciton state~\cite{Parish_EPL2011}. Recently, these results have been extended to the specific case of  TMDC monolayers~\cite{Cotlet_arxiv2018}.
Note that, in the presence of a Fermi sea, one cannot just consider the sign of $\exQdef$ to determine whether the many-body exciton state is bound or not. One must instead compare $\exQdef$ with the 
energy of the normal state, as defined next.

\subsection{Normal state} 
\label{sec:norma}
Under some conditions, we find that at large majority particle density, the finite CoM momentum exciton can undergo an unbinding transition 
to the normal (N) state.  This comprises an unbound minority particle on top of a Fermi sea of majority
particles:
\begin{equation}
  \ket{\Psi_{\text{N}}} = \hat{c}^\dag_{\0,2} \hat{c}^\dag_{\kf \hat{\k},1}
  \ket{\fs}   \; ,
\label{eq:norma}
\end{equation}
where $\hat{\k}$ is an arbitrary direction, and this state has energy 
\begin{equation}
  E_{\text{N}}=\ef - \frac{1}{\area} \sum_{\ve{k}'<\kf} U_{\kf \hat{\k} -
    \k'}\; ,
\label{eq:ennor}
\end{equation}
where, as for $E_{\Q}$, we are defining this with respect to $E_g$ and neglecting the energy of the interacting Fermi sea, $\mathscr{E}_{FS}$.

\begin{figure}
\includegraphics[scale=0.4]{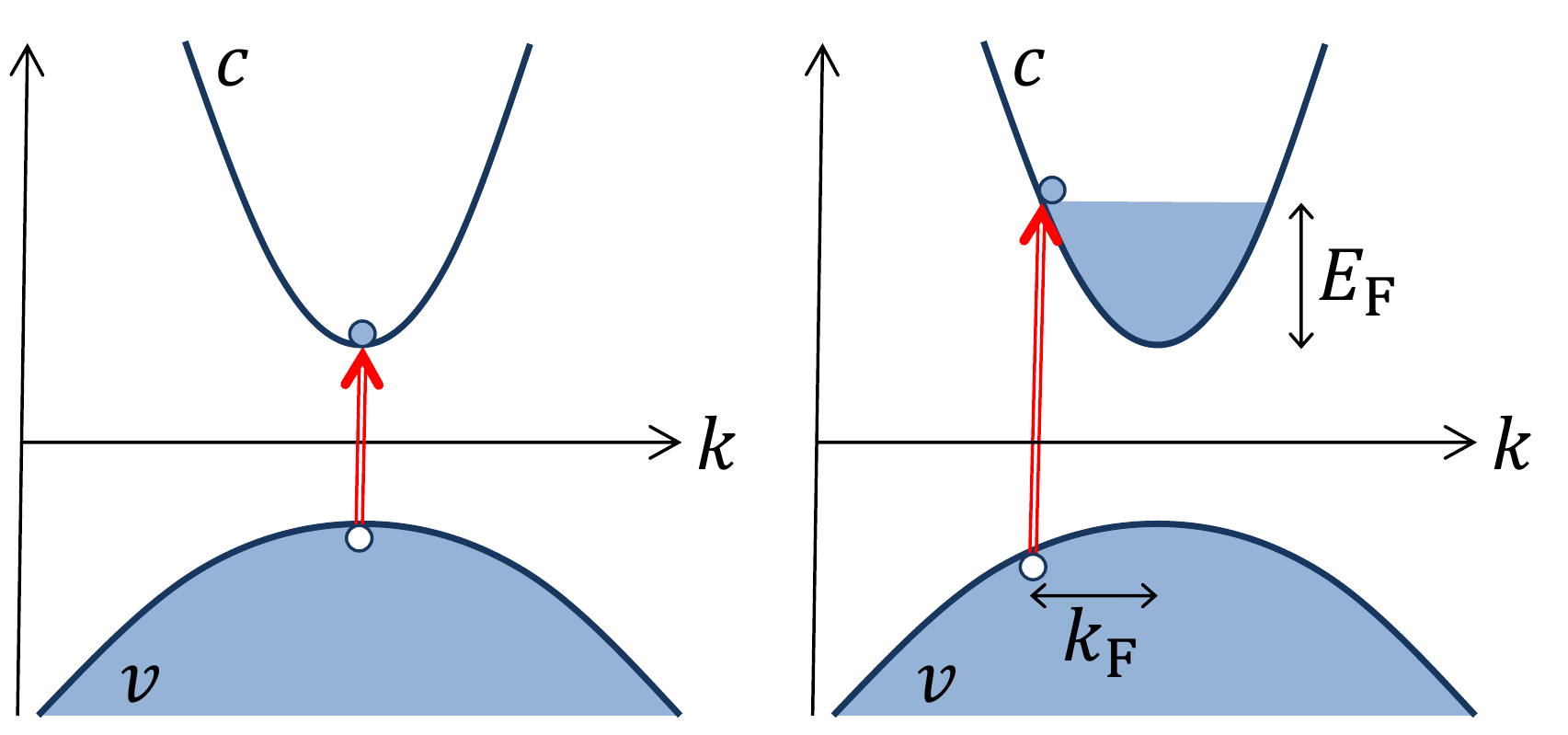}
\caption{Particle-hole excitation process via a photon with (a) and
  without (b) a Fermi sea --- all photon-mediated transitions are
  approximately vertical in a cavity.}
\label{fig:verti}
\end{figure}

In the absence of light-matter coupling, $g=0$, the excitonic FF state $\ket{\Psi_{\ve{Q}}}$~\eqref{eq:polar} would reduce to the normal state
$\ket{\Psi_{\text{N}}}$ when we take $\Q = \kf \hat{\k}$ and the exciton wave-function takes the form, $\varphi_{\ve{k}\ve{Q}} =
\sqrt{\area} 
\delta_{\k,\kf
  \hat{\k}}$. This corresponds to a wavefunction which has weight only when relative and CoM momenta are equal, and match the Fermi momentum $\kf$.

In Ref.~\onlinecite{Parish_EPL2011} it was shown that 
unscreened Coulomb interactions ($N_s=0$) always
lead to a bound many-body exciton state for any value of the density
and thus the normal state~\eqref{eq:norma} is never the ground state. We will show here that this is the case also in the presence of light-matter coupling.
When screening is non-zero however, a normal state can occur.
It is worth noting that when this state occurs, the only possible
normal state is purely electronic --- i.e., it has zero photon fraction 
and is thus given by Eq.~\eqref{eq:norma}.
This can be seen from the renormalization scheme of
the photon energy~\eqref{eq:renor}, which has the consequence that any non-zero photon fraction always implies a bound
state between minority and majority particles.
That is to say, the presence of light can bind an otherwise unbound electron-hole pair.

\subsection{Effective photon energy in the presence of a Fermi gas} 
\label{sec:effec}
In order to understand how the ground state evolves with doping, 
it is instructive to consider how the effective photon energy changes as the majority density increases, 
due to a modification of the dielectric constant of the quantum well.
As described in Sec.~\ref{sec:renor}, in order to reproduce the experimental protocol for measurements, we have defined the renormalization of the photon energy using a procedure defined at zero gating/doping $\ef=0$.  This means that we define the renormalized photon energy $\omega_{\text{C}\0}$ (or equivalently the photon-exciton detuning $\delta$) in such a way that it approximately matches what would be experimentally measured at $\ef=0$.
As illustrated in Fig.~\ref{fig:verti}, the available particle-hole excitations contributing to the dressing of the photon depend on $\ef$.
As such, at a finite density of majority species, the effective photon energy $\wCQef$ differs from $\omega_{\text{C}\vect{Q}}$ defined at $\ef=0$.
Here, we want to identify and estimate the magnitude of the photon energy shift in the presence of doping.

\begin{figure}
\centering
\includegraphics[width=1.0\columnwidth]{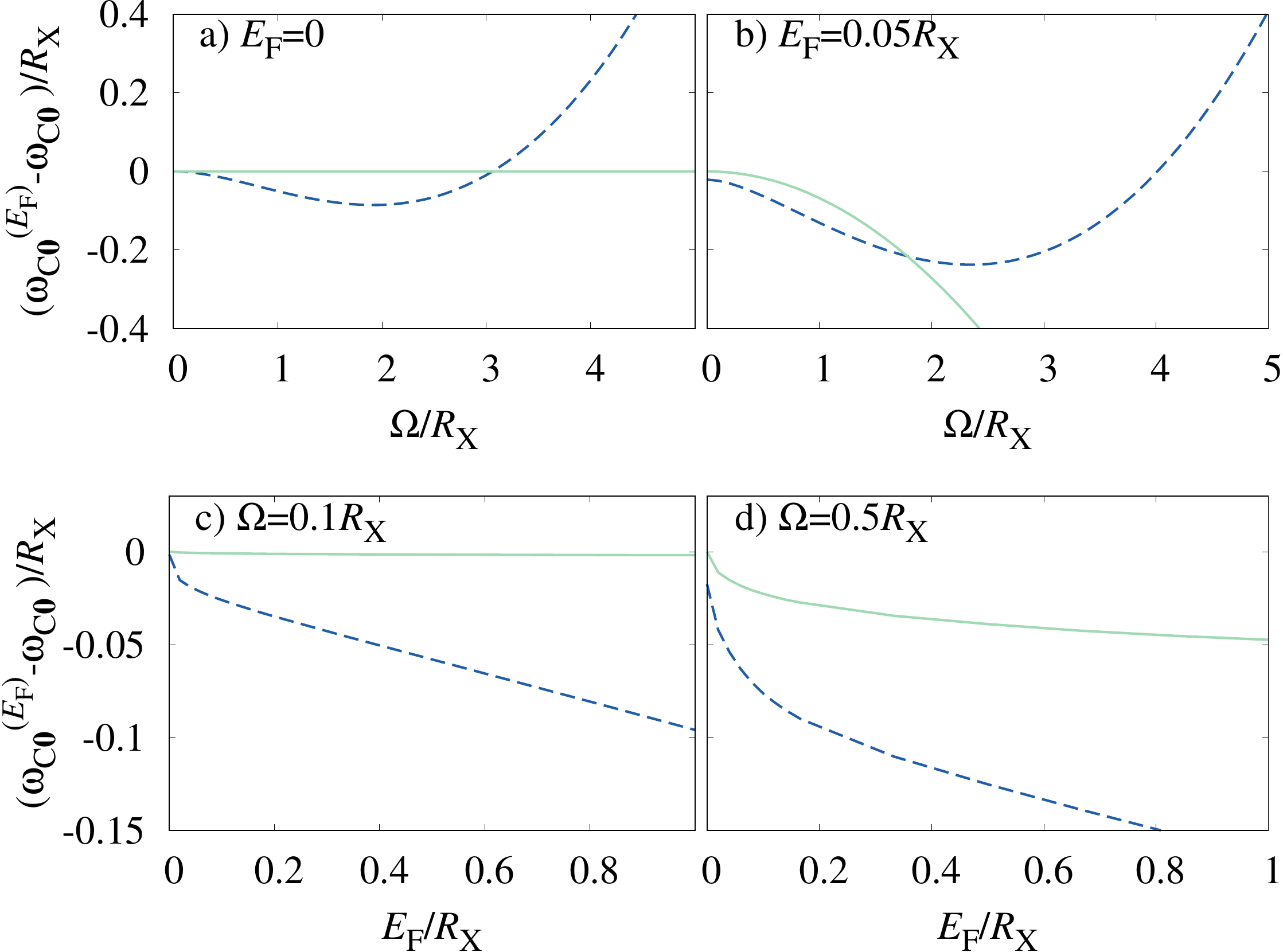}
\caption{Photon energy shift in presence of a Fermi gas $\wCQzeroef-\omega_{\text{C}\0}$ as estimated from Eq.~\eqref{eq:nudif} (solid line) and from Eq.~\eqref{eq:delfi} (dashed line) for either fixed Fermi energy $\ef$ and varying Rabi splitting $\Omega$ (top panels) or conversely fixed $\Omega$ and varying $\ef$ (bottom panels). Parameters are for a GaAs single quantum well ($d=0$), mass ratio $m_2/m_1=0.25$, and screened interactions $N_s=1$.}
\label{fig:phene}
\end{figure}
We start by rewriting the eigenvalue equations~\eqref{eq:eige1} and~\eqref{eq:eige2} in an equivalent form by inserting Eq.~\eqref{eq:eige1} in~\eqref{eq:eige2} and defining the new wavefunction $\beta_{\ve{k}\Q}= \frac{1}{\area}
\sum_{\ve{k}'>\kf}V_{\ve{k}-\ve{k}'}\varphi_{\ve{k}'\Q}/(-E_{\ve{Q}}+
\xi_{\k\Q})$:
\begin{multline}
\left(E_{\Q}-\nu_{\text{C}\vect{Q}}+E_g+\phantom{\Frac{g^2}{\area}
\sum_{\ve{k}>\kf}}\right.\\
\left. \Frac{g^2}{\area}
\sum_{\ve{k}>\kf}\frac{1}{-E_{\Q}+
  \xi_{\k\Q}}\right)\alpha_{\Q}=\frac{g}{\area}\sum_{\ve{k}>\kf}\beta_{\ve{k}\Q} \; .
\label{eq:rearr}
\end{multline}  
The divergence of the sum on the left-hand side of~Eq.~\eqref{eq:rearr} is exactly cancelled by the renormalization of the bare photon energy $\nu_{\text{C}\vect{Q}}$ by particle-hole excitations, as described in Sec.~\ref{sec:renor}.
The form of Eq.~\eqref{eq:rearr} suggests that, in the presence of a Fermi sea, the
effective renormalized photon energy can be estimated as
\begin{equation}
  \wCQef \simeq \nu_{\text{C}\Q} -
  \Frac{g^2}{\area} \sum_{\k>\kf} \Frac{1}{-E_{\ve{Q}}+\xi_{\k\Q}} \; .
\label{eq:effnu}
\end{equation}
This estimate is expected to be valid in the limit of small light-matter coupling and sufficiently small density, where there is a well-defined exciton bound state that is only weakly perturbed by light. In this limit, one can approximate $E_{\Q} \simeq \exQdef$.
Taking the CoM momentum to be zero, we then estimate the difference between $\wCQzeroef$ and the photon energy $\omega_{\text{C}\0}$ at zero doping \eqref{eq:renor} as
\begin{multline}
  \wCQzeroef - \omega_{\text{C}\0}
  \simeq- \Frac{g^2}{\area} \sum_{\k>\kf}
  \Frac{1}{-\exQzerodef + \xi_{\k\0}}\\
  +\Frac{g^2}{\area} \sum_{\k}
  \frac{1}{-\exd +\epsilon_{\k,1}+\epsilon_{\k,2}}\;.
\label{eq:nudif}
\end{multline}
This energy difference is clearly finite because the logarithmic divergence of the first sum cancels with the one of the second sum. Thus, we see that the photon energy shift with doping depends quadratically on the light-matter coupling strength $g$, provided $\Omega \ll |\exQzerodef -E_{\text{N}}|$. By numerically evaluating the density dependence of the exciton binding energy at $\Q=\0$, $\exQzerodef$ (see App.~\ref{app:phfra}), as well as the exchange correction to the electron dispersion, we find that, in the small $\Omega$ and $\ef$ limit, the photon energy shift $\wCQzeroef - \omega_{\text{C}\0}$ is always negative (see the solid line of Fig.~\ref{fig:phene}).
Such a shift could be observed in experiments by either comparing structures with different Rabi splittings or by changing the doping.

An alternative way of estimating the photon energy shift $\wCQzeroef - \omega_{\text{C}\0}$ in presence of a Fermi sea, is by identifying the detuning $\delta_{50\%}$ at which the many-body $\Q=\0$ exciton state and the cavity photon are at resonance:
\begin{equation}
   \wCQzeroef = \exQzerodef+E_g \; .
\label{eq:stron}   
\end{equation}
We assume that this condition is satisfied when the photon fraction $|\alpha_{\0}|^2$ is $1/2$.
We can rewrite the condition~\eqref{eq:stron}, which defines the detuning at resonance, $\delta_{50\%}$, by subtracting the energy of the photon mode at zero doping/gating  $\omega_{\text{C}\0}$~\eqref{eq:renor} from both sides. Then using the definition $\delta =  \omega_{\text{C}\0} - (\exd + E_g)$ on the right hand side gives:
\begin{equation}
    \wCQzeroef - \omega_{\text{C}\0} = \exQzerodef - \exd - \delta_{50\%} \; .
\label{eq:delfi}    
\end{equation}
We can thus estimate the photon shift $\wCQzeroef - \omega_{\text{C}\0}$ at a fixed value of $\ef$ and $\Omega$ by evaluating $\exQzerodef - \exd$, i.e., by solving Eq.~\eqref{eq:eige3}, and by numerically estimating the value of detuning $\delta_{50\%}$ at which the photon fraction is exactly $1/2$. The results of this estimate are plotted in Fig.~\ref{fig:phene} and compared with those obtained from Eq.~\eqref{eq:nudif}.
Note that, even at $\ef=0$, this estimate predicts a photon energy shift because, beyond the weak coupling regime $g \ll \ax \Ryx$, the exciton wavefunction is strongly modified by matter-light coupling, affecting the definition of detuning $\delta$ given in Eq.~\eqref{eq:detun} (see discussion in App.~\ref{app:fitti} and Fig.~\ref{fig:impro}).
At small and finite $\ef$, the estimates given by Eqs.~\eqref{eq:nudif} and~\eqref{eq:delfi} agree for small $\Omega$ giving a negative shift of the photon energy, while, when $\Omega$ increases, Eq.~\eqref{eq:delfi} predicts an upturn of the shift to positive values.

Predicting the exact behavior of $\wCQzeroef - \omega_{\text{C}\0}$ with either $\Omega$ or 
$\ef$ is non-trivial, since both estimates of Eqs.~\eqref{eq:nudif} and~\eqref{eq:delfi} are based on the assumption that the system does behave like a two-level coupled oscillator model, an hypothesis which looses validity when either $\Omega$ or $\ef$ increases.
As we will see in the next section, the shift of the photon energy with doping has little consequence for the phase diagram at fixed Rabi splitting $\Omega$, while the implications are larger when we fix $\ef$ and change $\Omega$.

\begin{figure*}
\centering
\includegraphics[width=1\textwidth]{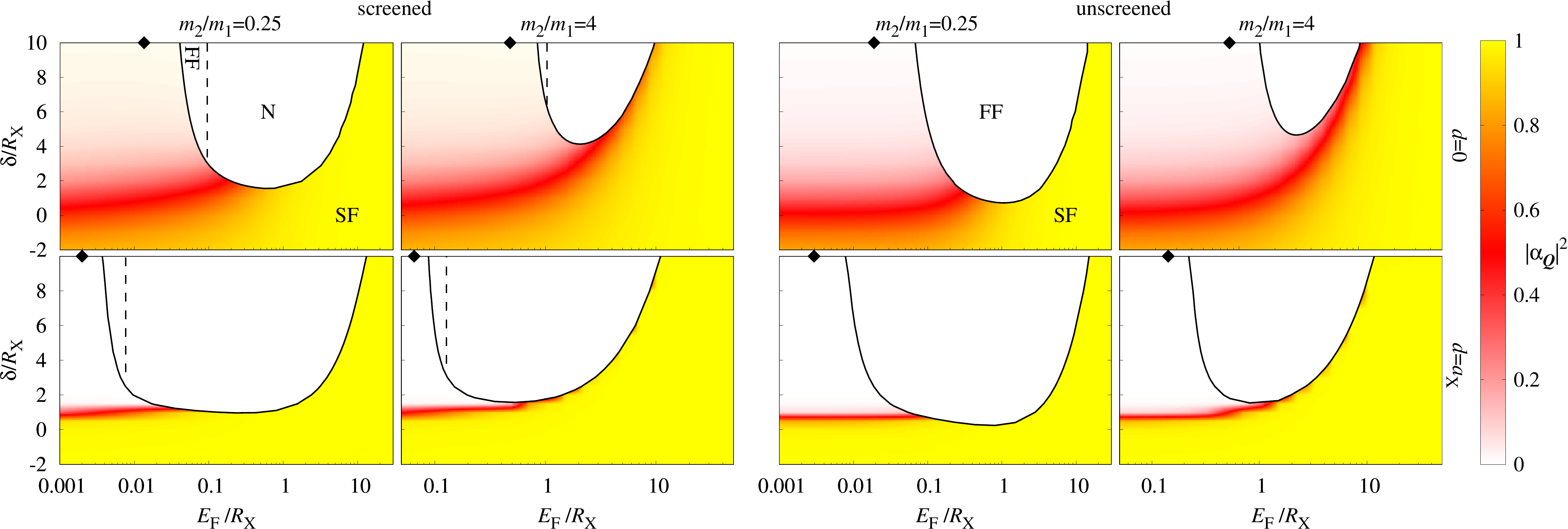}
\caption{Phase diagram of photon-exciton detuning $\delta$ and
  majority particle Fermi energy $\ef$ for a GaAs heterostructure with
  either a single quantum well ($d=0$, top panels) or a bilayer
  geometry at a distance $d=\ax$ (bottom panels). The
  four left panels are for RPA screened Coulomb interactions ($N_s=1$), while
  the right panels are for unscreened interactions ($N_s=0$). The mass ratio is
  fixed to either $m_2/m_1=0.25$, i.e., one electron in a Fermi sea of
  holes or $m_2/m_1=4$, i.e., one hole in a Fermi sea of electrons. The
  Rabi splitting is fixed to $\Omega = 2\Ryx$ for the $d=0$ case (top panels) and to $\Omega = 2|\exd|\simeq 0.64 \Ryx$ for the bilayer at $d=\ax$ case. Solid lines are 1$^{\text{st}}$-order
  transitions (SF-FF and SF-N). The dashed almost vertical line is the
  2$^{\text{nd}}$-order FF-N transition occurring for screened
  interactions. 1$^{\text{st}}$- and 2$^{\text{nd}}$-order transitions
  meet at a critical end-point. The diamond symbols indicate the value
  of the density, $\efzero$, at which the SF-FF transition occurs in the absence of
  the cavity field $\Omega=0=\alpha_\Q$. The color map represents the
  photon fraction $|\alpha_{\Q}|^2$.}
\label{fig:detun}
\end{figure*}
%
\subsection{Numerical implementation}
We obtain the ground-state phase diagram by numerically
diagonalizing the coupled equations~\eqref{eq:eigen} and analyzing the nature of the lowest energy state, while comparing it with the energy of the normal state~\eqref{eq:ennor}. We use a non-linear grid in the relative momentum
$\k$-space and evaluate, at a given value
of the CoM momentum $\Q$, the lowest eigenvalue $E_{\Q}$ and the associated excitonic $\varphi_{\k\Q}$ and
photonic $\alpha_{\Q}$ eigenvectors, with $|\alpha_{\Q}|^2$ representing the state photon fraction. The results we show are numerically converged with respect to the number of points employed in the momentum grid.
We then minimize the energy $E_{\Q}$ with respect to 
$Q\equiv|\Q|$, and
indicate the momentum at which the energy is minimized by
$Q_{\text{min}}$.

In the following we rescale
energies by the 2D exciton binding energy $\Ryx$ and lengths by the
Bohr radius $\ax$ defined in  Eqs.~\eqref{eq:ebohr}. Hence, only a few independent dimensionless
parameters are left to characterize the system properties and phase diagram, namely, the mass ratio between minority and majority particles $m_2/m_1$, the rescaled
bilayer distance $d/\ax$, the dimensionless majority particle density 
$\ef/\Ryx$, the photon-exciton detuning $\delta/\Ryx$, Eq.~\eqref{eq:detun},
and the Rabi splitting $\Omega/\Ryx$, Eq.~\eqref{eq:rabic}.

\section{Charge-imbalanced quantum wells in planar microcavities}
\label{sec:resul}
We first consider the case of a GaAs quantum well system embedded in a microcavity. In Fig.~\ref{fig:detun} we show our calculated phase diagram as a function of majority particle density and detuning, keeping the Rabi splitting fixed. We compare the results for both screened and unscreened Coulomb interactions, for a single quantum well ($d=0$) and a bilayer geometry ($d=\ax$), and for one electron in a Fermi sea of holes ($m_2/m_1=0.25$) and one hole in a Fermi sea of electrons ($m_2/m_1=4$). In all cases, we see that the coupling to cavity light modes
suppresses the formation of the finite momentum FF state
as compared to the case without light-matter coupling. 
In particular, a strong coupling to light favors the $\Q=\0$ state, since the photon mode at non-zero $\Q$ is at high energy, due to the small photon mass. As such, strong coupling to light imposes that for detunings below a minimal value, $\delta<\delta_{\text{min}}$, only the $\Q=\0$ SF phase is allowed.

\begin{figure}
\centering
\includegraphics[width=1\columnwidth]{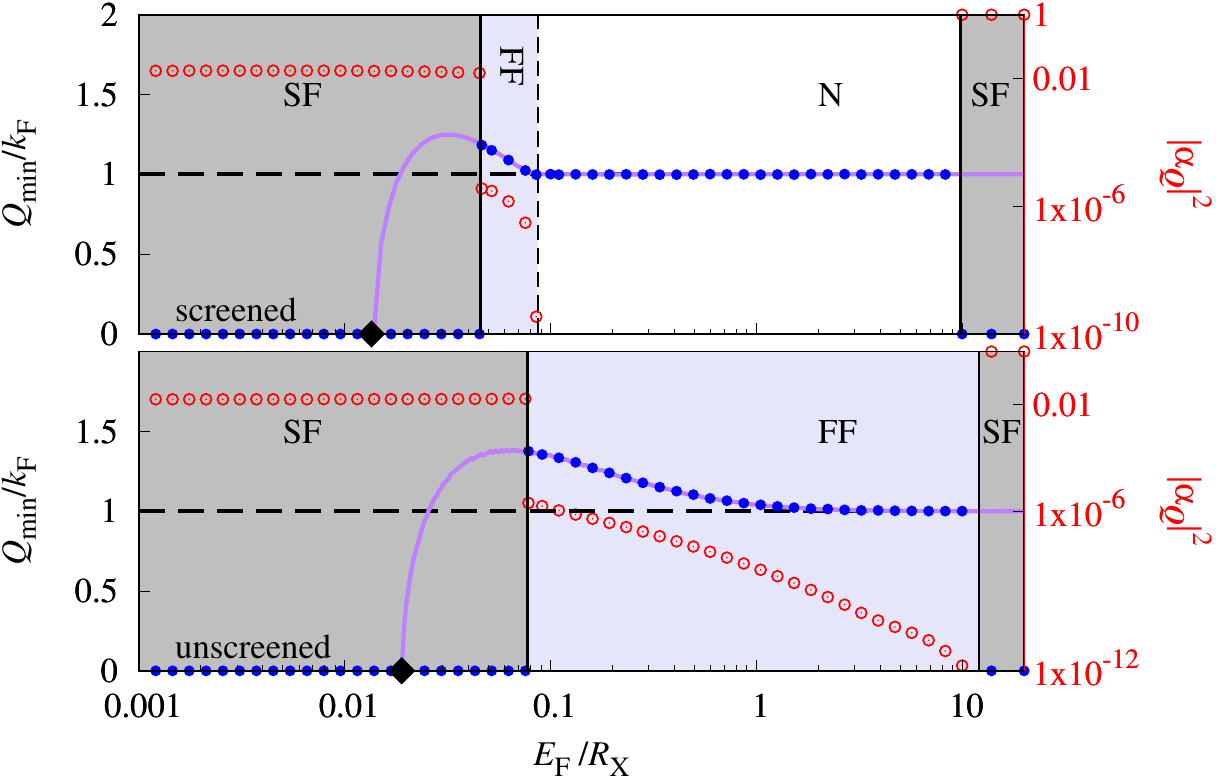}
\caption{ Momentum $Q_{\text{min}}$ (filled [blue] circles) minimizing the
  many-body polaritonic energy $E_{\Q}$ as a function of the 
  majority Fermi energy $\ef$ for a single quantum well $d=0$, mass ratio
  $m_2/m_1=0.25$, Rabi splitting $\Omega = 2\Ryx$ and detuning
  $\delta=8\Ryx$. Interactions are RPA screened ($N_s=1$) in the top panel, and unscreened ($N_s=0$) in the bottom
  panel. Solid (violet)
  lines represent the value of $Q_{\text{min}}$ in the absence of
  light-matter coupling ($\Omega = 0$), while the thick dashed (black)
  line is the Fermi momentum $\kf$. The corresponding photon fraction
  $|\alpha_{\Q}|^2$ is plotted with open (red) circles and the
  corresponding axes are on the right side of each panel.}
\label{fig:qminf}
\end{figure}

Fixing the detuning $\delta>\delta_{\text{min}}$ and increasing $\ef$, one first finds a SF-FF transition between a $\Q=\0$ many-body mixed polariton state and a $Q_{\text{min}}\ne 0$ FF state weakly coupled to light. 
This state has also been referred to as a roton minimum~\cite{Cotlet_arxiv2018}.
This occurs because the energy gained by forming a finite $\Q$ exciton state is larger than that obtained by
dressing the $\Q=\0$ exciton with a zero momentum photon.
For screened interactions the transition can be directly to the unbound N state, while for unscreened interactions there is no normal phase, just as in the absence of photons~\cite{Parish_EPL2011}.
Both the SF-FF and SF-N transitions are first order (see App.~\ref{app:order}), with $Q_{\text{min}}$ changing discontinuously from $Q_{\text{min}}=0$ to a finite value, as shown in Fig.~\ref{fig:qminf}.
Because of the small cavity photon mass, the finite $Q_{\text{min}}$ FF phase has a small photon fraction, that decreases further on increasing $\ef$ (see Fig.~\ref{fig:qminf}). Thus, the value of $Q_{\text{min}}$ almost coincides with that in the absence of the cavity field, and in particular $Q_{min}$ locks to $\kf$ at the FF-N transition. In contrast, for unscreened interactions, $Q_{min}$ asymptotically tends to $\kf$ in the FF region only for large values of $\ef$.
In addition, the FF-N transition is always second order and it is only weakly affected by the coupling to light --- thus it is approximately independent of both $\delta$ and $\Omega$.

The SF-FF transition is strongly affected by the coupling to a cavity field. In particular, the many-body exciton at $\Q=\0$ strongly couples to the cavity photon when both energies are comparable, resulting in a half-matter half-light many-body polariton state. In Fig.~\ref{fig:detun}, the red region of the color map indicates where the photon fraction is around $50\%$, corresponding to resonance between the cavity photon and the many-body exciton. The value of the detuning $\delta$ for which resonance occurs is seen to grow with the majority density.
This is mostly due to the $\Q=\0$ exciton energy $\exQzerodef$ growing with $\ef$ due to Pauli blocking (see App.~\ref{app:phfra}). Indeed, one can show that $\exQzerodef$ grows sub-linearly for $\ef \ll \Ryx$ and screened interaction, while it grows linearly $\sim \ef$ for $\ef>\Ryx$ (see App.~\ref{app:phfra} and Figs.~\ref{fig:exenr} and~\ref{fig:phfra}).

At large positive detunings, we recover, as expected, the results obtained in Ref.~\onlinecite{Parish_EPL2011} for GaAs single wells and bilayers in the absence of light-matter coupling. %
Here, as one increases the majority particle density, Pauli blocking causes the many-body exciton energy $E_{\Q}=\exQdef$ obtained by solving Eq.~\eqref{eq:eige3}  to develop a minimum at finite CoM momentum $Q_{\text{min}}$, as this reduces the kinetic energy cost of the minority particle. 
We denote the Fermi energy 
at which this transition occurs in the excitonic limit by $\efzero$, and, in the figures, this is illustrated by a diamond symbol. Without light, the transition to the FF state is always second order (see App.~\ref{app:order}).

\begin{figure*}
\centering
\includegraphics[width=0.8\textwidth]{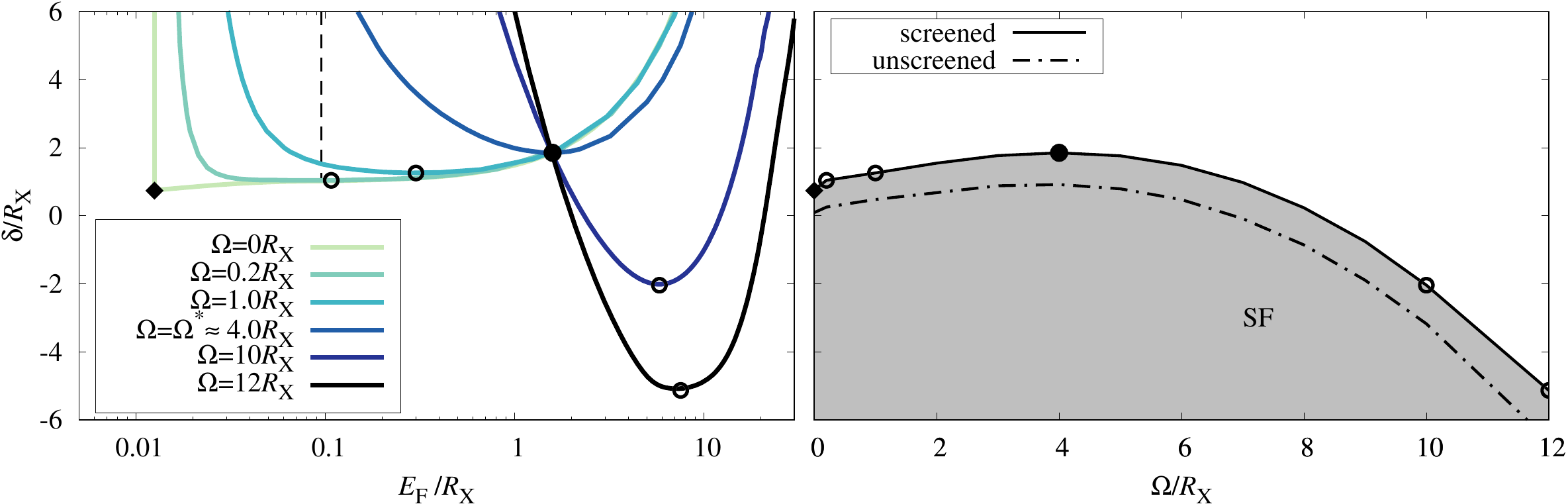}
\caption{Left panel: Solid lines are SF-FF (or SF-N) phase boundaries for different values of the Rabi splitting $\Omega$, for a single quantum well
  with hole doping, $d=0$ and $m_2/m_1=0.25$, and for screened interactions $N_s=1$. In particular, the region above
  a solid line is either FF (on the left of the dashed line) or N (on the right). The almost vertical dashed line is the approximately $\Omega$-independent FF-N
  boundary (see Fig.~\ref{fig:detun}). Below each solid line the phase is SF.
  Each symbol represents the minimal
  detunings $\delta_{\text{min}}$ of the boundaries --- special values
  are $\Omega=0$ (filled diamond) and the $\Omega$ at which $\delta_{min}=\delta^*$
  (filled circle). A special value common to all boundaries is $(\delta^*,\ef^*)$ (filled circle).  Right panel:  the solid line and  symbols give the behavior of
  $\delta_{\text{min}}$ as a function of $\Omega$ for screened $N_s=1$ interactions, while the dot-dashed line represents $\delta_{\text{min}}$ for unscreened $N_s=0$ interactions.}
\label{fig:chang}
\end{figure*}

By further increasing  the density at fixed (large positive) photon-exciton detuning, there is eventually an additional first order transition to an almost completely photon-like $\Q=\0$ SF state. This is because the energy of the FF and N states is pushed up by Pauli blocking such that they exceed the photon energy at sufficiently large density. 
As such, larger values of the detuning require larger values of density for this second transition to occur. 
Since this transition only weakly depends on the light-matter coupling,  
the FF-SF (N-SF) boundary essentially occurs when $\delta \simeq \exQdefmin  - \exd$ ($\delta \simeq E_{\text{N}} - \exd$), where  $\exQdefmin$ is the FF many-body exciton energy at Fermi energy $\ef$ and bilayer distance $d$ in the absence of the photon field -- see Eq.~\eqref{eq:eige3}.

From the study of the phase diagram at fixed Rabi splitting, we can draw similar conclusions about the mechanisms promoting the existence of a FF phase to those known in the absence of the cavity photon~\cite{Parish_EPL2011}: the FF phase is favored by 
unscreened Coulomb interactions and by a small minority particle mass. 
In addition, considering the unscreened case, a finite bilayer distance also  favors FF. This is because the inter-layer interaction suppresses large momentum scattering and promotes an exciton wave-function $\varphi_{\k\Q}$ peaked at the $\k\sim \Q$ direction, and also because a finite inter-layer distance reduces the effective electron-hole coupling to light.
While our results demonstrate that embedding the quantum well structure into a cavity  reduces the parameter region where FF can occur, 
this phase is still weakly coupled  to light. 
Thus, the FF ground state should be visible in the photon momentum distribution, in an experiment with sufficient sensitivity.
Note that for our simplified scenario in Eq.~\eqref{eq:polar}
of a single minority particle and thus a single photon in the cavity, the system
photoluminescence is 
peaked at the energy 
$E_{\Q}+E_g$, 
with a weight given by the corresponding photon fraction $|\alpha_{\Q}|^2$.
\begin{figure*}
\centering
\includegraphics[width=1\textwidth]{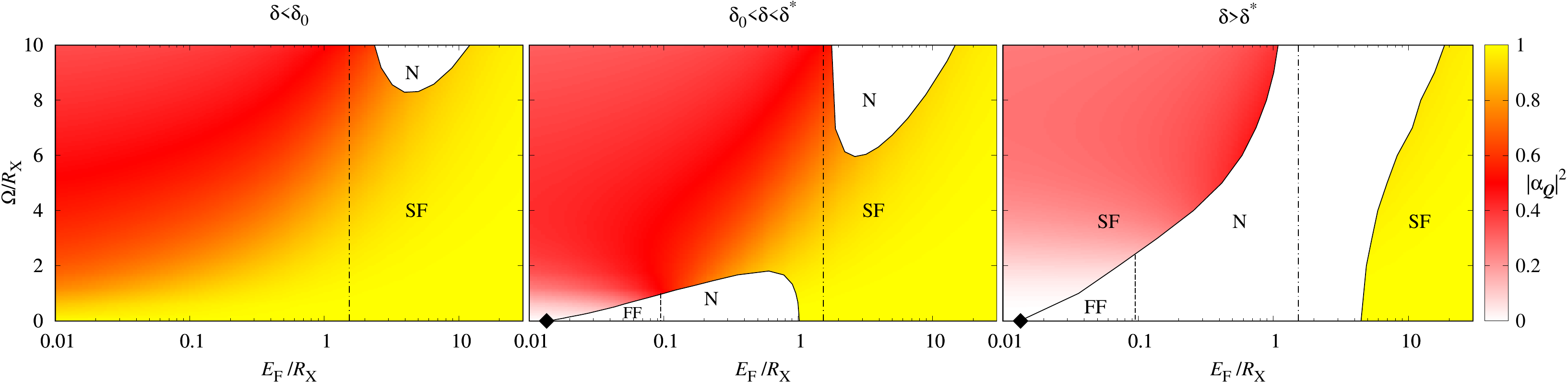}
\caption{Different possible topologies of the phase diagram as a
  function of the of the Rabi splitting $\Omega$ and the majority
  particle Fermi energy $\ef$ for a GaAs heterostructure with a single
  quantum well ($d=0$), $m_2/m_1=0.25$, and screened interactions $N_s=1$. The detuning has been
  fixed to $\delta=0<\delta_0$ (left panel), $\delta_0 < \delta=1.5
  \Ryx<\delta^*$ (middle panel), and $\delta=4\Ryx>\delta^*$ (right
  panel). In all panels, the vertical dot-dashed line is the value of
  $\ef^*$ (see Fig.~\ref{fig:chang}), while all other lines and labels are as in Fig.~\ref{fig:detun}.}
\label{fig:omega}
\end{figure*}
%
\subsection{Comparison of structures with different Rabi splitting}
\label{sec:compa}
It is possible to study the evolution of the FF phase with changing Rabi splitting by considering a sequence of cavities which have different numbers $N_{QW}$ of embedded quantum wells, since $\Omega\sim\sqrt{N_{QW}}$\cite{Weisbuch1992,Bloch_APL_1998}.
In particular, in Ref.~\onlinecite{Brodbeck_PRL2017}, two structures with either 1 or 28 quantum wells stacked at the antinodes of the cavity field have been compared, allowing one to study the change of the Rabi splitting in the range $0.3\Ryx\lesssim\Omega\lesssim 1.3 \Ryx$.
Note that in inorganic microcavities, while typically $\Omega \ll E_g $, the very strong coupling regime $\Omega > \Ryx$ can also be routinely achieved~\cite{Bloch_APL_1998,Saba_Nature2001,Kasprzak_Nature2006}. Studying the evolution of the phase diagram with increasing Rabi splitting should in principle directly show how the introduction of light-matter coupling modifies the phase diagram.

With this motivation, in the left panel of Fig.~\ref{fig:chang}, we compare  the boundaries between the SF and the FF (SF and N) phases for different values of $\Omega$. Screened and unscreened interactions give qualitatively the same results, with the only difference being the absence of the N phase for unscreened interactions. The boundaries are also quantitatively similar in the two cases.
In the absence of light-matter coupling, the SF-FF boundary is given by ($\ef>\efzero$):
\begin{equation}
  \delta = \exQdefmin- \exd\; ,
\label{eq:Omeg0}  
\end{equation}
where $\exQdefmin$ is the FF many-body exciton energy at Fermi energy $\ef$. 
For the SF-N boundary at $\Omega=0$, this expression becomes $\delta = E_{\text{N}}- \exd$.
We observe an evolution of the minimal photon-exciton detuning $\delta_{\text{min}}$ with $\Omega$ (right panel of Fig.~\ref{fig:chang}) which, starting from the value  $\delta_0 = E_{\text{X}\Q_{\text{min}}}^{(d,\efzero)} - \exd$ at $\Omega=0$, grows up to a maximum value $\delta^*$, and then decreases again.
Consequently, 
the light-matter coupling is detrimental to the formation of a finite momentum phase for small values of $\Omega$, while it favors finite $\Q$ at $\Omega \gtrsim 4\Ryx$.

There is a special point  $(\delta^*,\ef^*)$ which is common to all SF-FF (SF-N) boundaries as one varies $\Omega$, i.e., one observes in the left panel of Fig.~\ref{fig:chang} that all lines appear to cross at a single point.
At this particular value of the photon-exciton detuning and density, all the dependence on the Rabi splitting and thus the light-matter coupling is lost.
Here, the decrease in energy due to forming a polariton is exactly counterbalanced by doping-induced changes to the cavity dielectric constant discussed in Sec.~\ref{sec:effec}. Note that this behavior is not accurately captured by the estimated photon shift in Eq.~\eqref{eq:nudif}, since this is not valid in the regime $\ef > \Ryx$. However, we can determine $(\delta^*,\ef^*)$ once we account for all the electron-hole scattering processes, as shown in App.~\ref{app:efsta}.
We have checked that the existence of the special point $(\delta^*,\ef^*)$ is common to both structures with single well and bilayer geometry, and it is also independent of whether interactions are screened or unscreened.

To further illustrate the special role played by the detuning $\delta^*$ and Fermi energy $\ef^*$, we plot in Fig.~\ref{fig:omega} the three different types of phase diagrams at fixed detuning $\delta$ that arise by varying $\Omega$ and $\ef$. 
A common feature for all three cases is that, for $\ef<\ef^*$, the FF and N phases are suppressed on increasing $\Omega$, in favor of a strongly mixed light-matter polaritonic SF phase with $|\alpha_{\0}|^2 \sim 0.5$. Note also that for $\ef<\ef^*$ the FF (N) phase occurs only for $\delta>\delta_0$.
In this small $\ef$ case, the lowering of energy of the strongly mixed $\Q=\0$ LP state with $\Omega$ dominates over any change of the cavity dielectric constant 
 because of gating/doping.
Note that the phase diagram we see in this small $\ef$ case illustrates the idea that increasing light-matter coupling can stabilize a polaritonic ground state even when the purely excitonic system is unbound.

For $\ef>\ef^*$, we see quite a different behavior --- a finite momentum FF or N phase is favored at larger values of the Rabi splitting $\Omega$, regardless of the value of the detuning.
In this large $\ef$ case, the SF-FF (SF-N) transition typically occurs from an almost purely photonic SF phase $|\alpha_{\0}|^2 \sim 1$ to an almost purely excitonic FF (N) phase with $|\alpha_{\0}|^2 \ll 1$ ($|\alpha_{\0}|^2 = 0$). This transition occurs because the shift in the cavity dielectric constant 
at finite $\ef$ increases with $\Omega$, while the excitonic or normal state energy is $\Omega$ independent, so that eventually, increasing $\Omega$ to large enough values, one favors the excitonic phase over the polaritonic.

Note that for GaAs heterostructures with a single quantum well and $m_2/m_1 = 0.25$, we find that $\ef^*\simeq 1.55 \Ryx$ ($\ef^*\simeq 1.95 \Ryx$) for screened $N_s=1$ (unscreened $N_s=0$) interactions respectively --- see App.~\ref{app:efsta}. This value of the Fermi energy is well below typical energies at which band curvature and structure start being important, so it lies within the range of the validity of our model. Indeed, from the GaAs lattice constant $a\simeq 0.56$~nm, we can estimate that 
$1/(2\mu a^2) \simeq 150 \Ryx \gg \ef^*$.

\section{TMDC monolayer embedded into a planar microcavity}
\label{sec:tmdcv}
\begin{figure}
  \includegraphics[width=0.9\columnwidth]{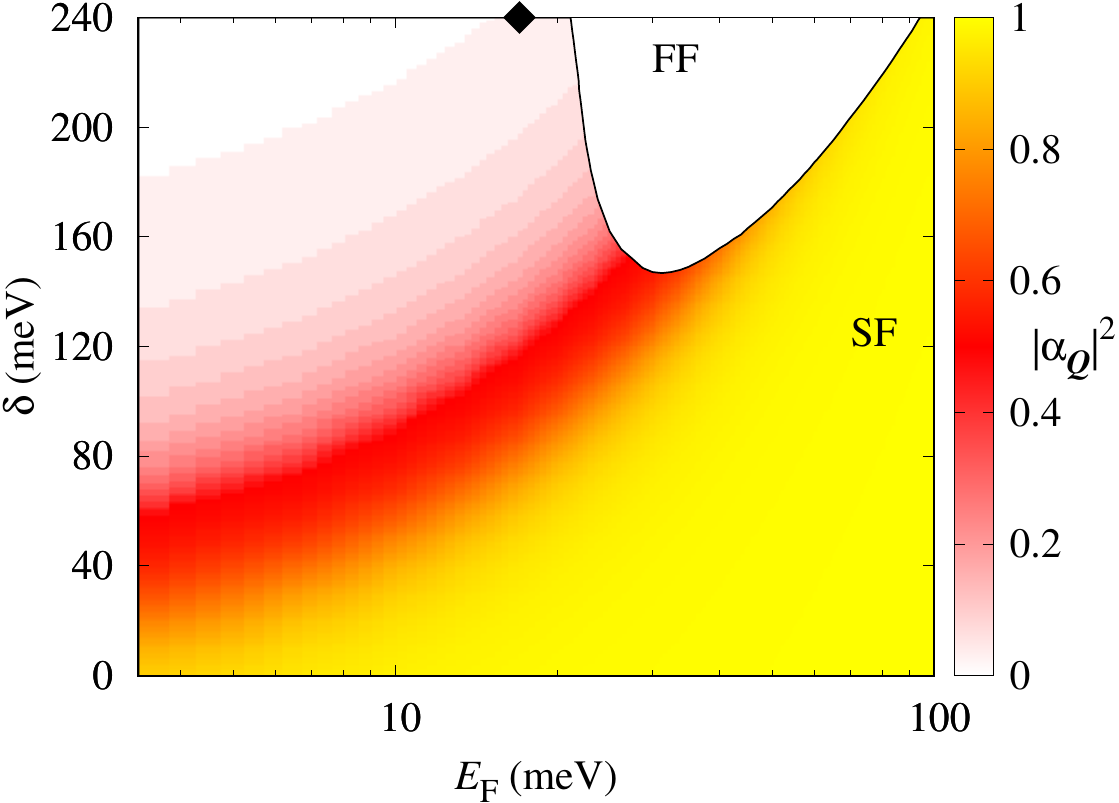}
  \caption{Phase diagram for a MoSe$_2$ monolayer embedded into a
    planar cavity as a function of  photon-exciton detuning $\delta$ and
    electron Fermi energy $\ef$. We take the Rabi splitting $\Omega = 40 \text{meV}$. The
    solid line is the 1$^{\text{st}}$-order SF-FF transition, while
    the diamond symbol indicates the value of $\efzero$ at which the
    SF-FF transition occurs in the absence of the cavity field
    ($\Omega=0=\alpha_\Q$)~\cite{Cotlet_arxiv2018}. The color map represents the
    photon fraction $|\alpha_{\Q}|^2$.}
\label{fig:tmdcp}
\end{figure}

As mentioned in the introduction, one context in which electronically doped polariton systems have been studied experimentally are TMDC materials~\cite{Sidler_NP2016,Chen_NaturePh_2017,Zheng_NaturePh_2017,Liu_PRL_2017}. 
We derive here the phase diagram for the specific case of doped MoSe$_2$, see Fig.~\ref{fig:tmdcp}. In particular, we consider the case of a single hole in a Fermi sea of electrons, with all electrons being spin and valley polarized, a regime which can be experimentally realized by applying a magnetic field~\cite{Back_PRL_2017}.

Due to the fact that most of the dielectric screening takes place within the two-dimensional layer, TMDC materials require a separate analysis from the case of III-V semiconductor heterostructures.
Specifically, we consider the same model
Hamiltonian as before, Eq.~\eqref{eq:hamge}, with a screened electron-hole interaction appropriate for  a monolayer in vacuum~\cite{Keldysh_1979,Rytova,Hanan2018}:
\begin{equation}
   V_{\ve{q}}^{\text{RK}} = \frac{2\pi
    e^2}{q}\frac{1}{\left( 1+r_0 q \right)}\; .
\label{eq:keldy}    
\end{equation}
For MoSe$_2$, the screening length is $r_0 = 5$~nm~\cite{Berkelbach_PRB2013}.
Note that, in contrast to Thomas-Fermi screening, the dielectric screening 
vanishes at large distances,
i.e., $V_{\q}^{\text{RK}} \to 2\pi e^2/q$ for $q\to 0$.
The electron and hole masses are $m_1 \equiv m_e = 0.56 m_0$ and $m_2 \equiv m_h = 0.59 m_0$~\cite{Berkelbach_PRB2013,Rasmussen2015}, where $m_0$ is the free electron mass. Because $m_e$ and $m_h$ have very similar values, little difference is expected whether the minority species is a hole --- as explicitly considered here --- or an electron. 
Following Ref.~\onlinecite{Cotlet_arxiv2018}, we neglect 
electron exchange; furthermore,  we neglect 
screening by the electron gas on the basis that, for these materials, the plasma frequency, $\omega_{\text{pl}} (\kf) \simeq \sqrt{n_1 \kf^2 V_{\kf}^{\text{RK}}/m_1} \sim 90$~meV (for $\ef=20$~meV), is much smaller that the exciton binding energy $|\ex|=485$~meV~\cite{Berkelbach_PRB2013,Ugeda_NatMat2014,Rasmussen_JPCC2015}.

For TMDC monolayers, strong coupling to light can be attained by placing flakes of material in planar cavities. Strong light-matter coupling leading to exciton-polariton formation is by now routinely achieved~\cite{Menon_NatPh_2014,Dufferwiel2015,Chen_NaturePh_2017,dufferwiel2017valley,sun2017optical}. We fix the cavity photon mass to $m_{\text{C}}=10^{-5} m_0$ and the Rabi splitting to $\Omega=40$~meV~\cite{Menon_NatPh_2014}. Note that, even with such a large value of $\Omega$,  because  the exciton binding energy 
is even larger, the regime of very strong coupling~\cite{khurgin2001excitonic} $\Omega \gtrsim |\ex|$ has not yet been reached for TMDCs. However, recently, there has been strong progress 
in this direction, see, e.g., Refs.~\onlinecite{schneider2018two,waldherr2018observation}.
Importantly for our analysis, the renormalization scheme of the photon energy described in Sec.~\ref{sec:renor} is unchanged. 

By considering the same variational many-body polariton state as in Eq.~\eqref{eq:polar} we derive the phase diagram versus detuning $\delta$ and electron Fermi energy 
$\ef$. The resulting phase diagram is shown in Fig.~\ref{fig:tmdcp}, and is seen to qualitatively agree with the unscreened case of GaAs presented in Fig.~\ref{fig:detun}.
Because the long-range unscreened Coulomb interaction promotes the finite momentum bound FF phase, it is not surprising that the system never transitions to the normal state N for the potential in Eq.~\eqref{eq:keldy}.  As shown in Ref.~\onlinecite{Parish_EPL2011}, the bare Coulomb interaction always implies a bound exciton state for any density of majority particles. %
In the absence of the cavity photon mode, we recover the results of Ref.~\onlinecite{Cotlet_arxiv2018}, which predicted a SF-FF transition at $\efzero = 20$~meV--- as before, this value is labelled with a diamond symbol in Fig.~\ref{fig:tmdcp}.
Because of the large value of $|\ex|$ relative to $\Omega$, the minimal photon-exciton detuning for observing FF is found to be rather large, $\delta_{\text{min}} \simeq 147$~meV. However, we expect this value to eventually decrease for $\Omega\gtrsim|\ex|$ in a manner similar to that shown in Fig.~\ref{fig:chang}.

\section{Conclusions and perspectives}
\label{sec:conc}
We have studied polaritonic phases in an extremely charge imbalanced electron-hole mixture in either a single quantum well, a bilayer or for TMDC monolayers embedded into a planar cavity.
In particular, we have analysed the competition between the formation of an FF-like~\cite{Fulde-Ferrell_PR1964} bound excitonic pair at finite CoM momentum, which is promoted by both long-range Coulomb interactions and the Pauli blocking of the Fermi sea~\cite{Parish_EPL2011,Cotlet_arxiv2018}, and 
 the formation of a strongly coupled many-body polariton state at zero momentum, which is promoted by
the strong coupling to the cavity field.
By fixing the light-matter coupling, i.e., the Rabi splitting, we find that, as expected,  strong coupling to a cavity photon mode competes against the formation of the finite momentum FF state, and so reduces the  parameter range of majority species density where this phase occurs.
Note that the the FF phase does weakly couple to light so that to allow its detection in photoluminescence experiments with enough sensitivity.
For large  photon-exciton detunings the photon becomes less relevant, and so the FF phase occupies a sizeable region at finite density of the majority species. At small densities the FF phase is replaced by bound polariton states with zero CoM momentum,  which lower their energy through strong light-matter coupling. At large densities, one instead finds an almost purely photonic state (with zero momentum) because, due to Pauli blocking, the exciton energy grows roughly linearly with the density.  As already known for the case without photons, a bound state always exists for unscreened Coulomb interactions, whereas with screening, an unbound state can replace the excitonic FF state.

To understand the topology of the phase diagram, we note that it is important that the presence of a Fermi sea not only changes the energy of the exciton but also the background cavity dielectric constant 
of the active medium, i.e., the gated/doped quantum well, the bilayer or the TMDC monolayer.
This change has little consequences for the phase diagram at fixed Rabi splitting, because the exciton energy shift with density dominates over the shift of the photon energy.
However, the photon energy shift increases for sufficiently large values of the Rabi splitting, and consequently does have a significant effect on the phase diagram at fixed detuning.
In particular, we find that increasing the Rabi splitting at low enough doping/gating densities always promotes the formation of a zero momentum strongly bound polariton state. However, surprisingly, at large enough densities, this behavior is reversed and increasing the coupling to light promotes the formation of finite momentum excitonic states weakly mixed to light.

The results in this paper focus entirely on the regime of extreme imbalance, where there is only a single minority species particle.  It is of course interesting to consider the behavior of the many-body state with a larger minority particle density;  this will be discussed in a subsequent paper~\cite{strashko2019}. Another important question concerns the possibility of more complex pairing states, even in the extreme imbalance state.  The Ansatz we use in this paper assumes that the pairing state has no effect on the majority Fermi sea, however Coulomb interactions between majority particles mean this assumption will not necessarily hold.  Relaxing this assumption allows the excitonic state to be dressed by electron-hole pairs of the majority band --- such effects have been considered recently for a tightly bound exciton in doped TMDCs~\cite{Sidler_NP2016,Efimkin_PRB2017}.
Understanding the interplay of this dressing with the internal structure of pairing, the coupling to light, and the crossover from the behavior we discuss here to the Fermi-edge polariton regime is a topic for future work.

\acknowledgments 
AT and FMM acknowledge financial support from the Ministerio de
Econom\'ia y Competitividad (MINECO), project No.~MAT2017-83772-R.  JL
and MMP acknowledge support from the Australian Research Council
Centre of Excellence in Future Low-Energy Electronics Technologies
(CE170100039).  JL is also supported through the Australian Research
Council Future Fellowship FT160100244.
AHM and JK acknowledge financial support from a Royal Society International Exchange Award, IES\textbackslash{}R2\textbackslash{}170213.
JK acknowledges financial
support from EPSRC program ``Hybrid Polaritonics'' (EP/M025330/1).
AHM acknowledges support from Army Research Office (ARO) Grant \# W911NF-17-1-0312 (MURI) by Welch Foundation Grant TBF1473. 
This work was performed in part at Aspen Center for Physics, which is supported by National Science Foundation grant PHY-1607611. This work was partially supported by a grant from the Simons Foundation.
%
\appendix

\begin{figure}
\centering
\includegraphics[width=1.0\columnwidth]{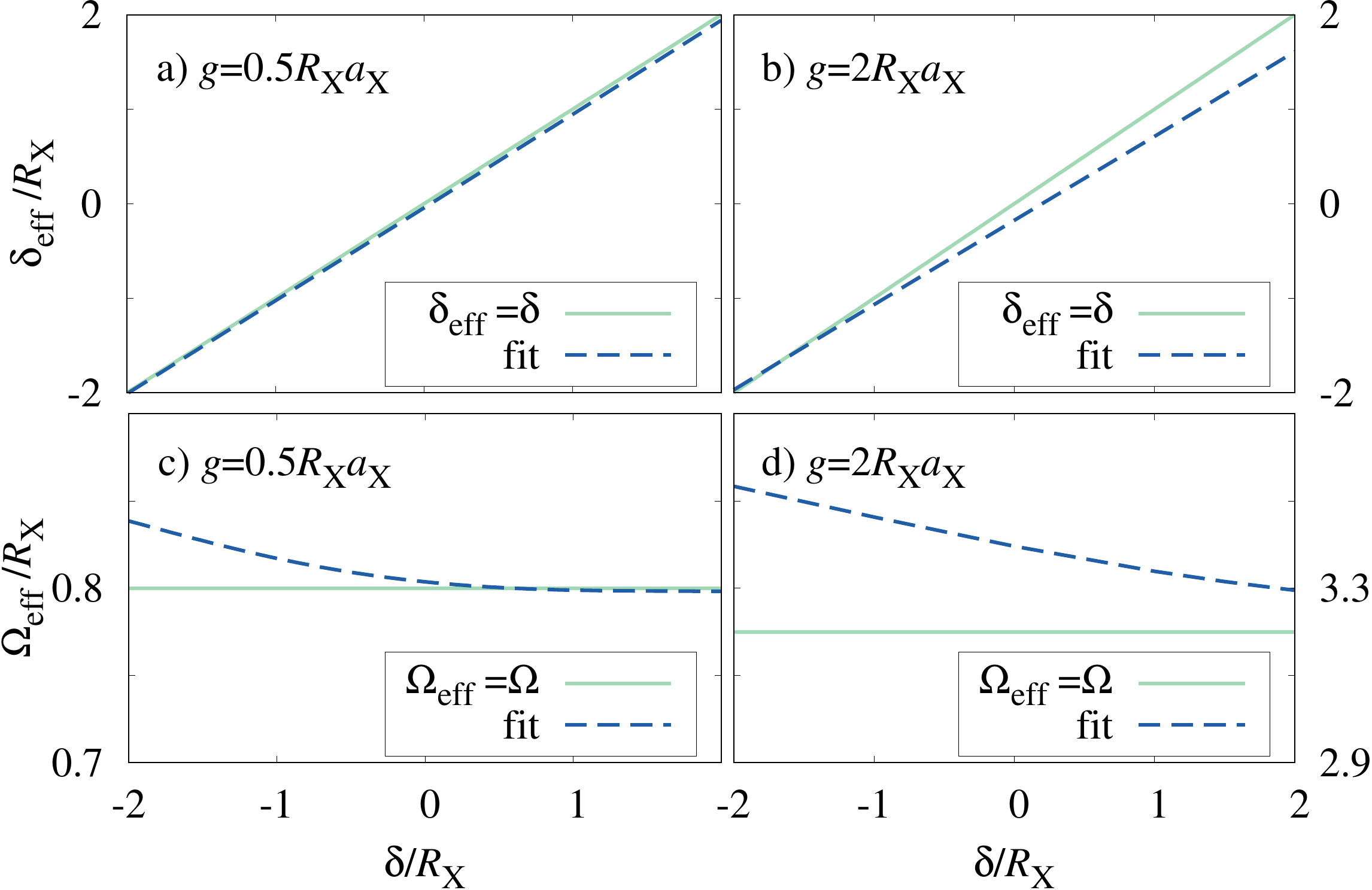}
\includegraphics[width=1.0\columnwidth]{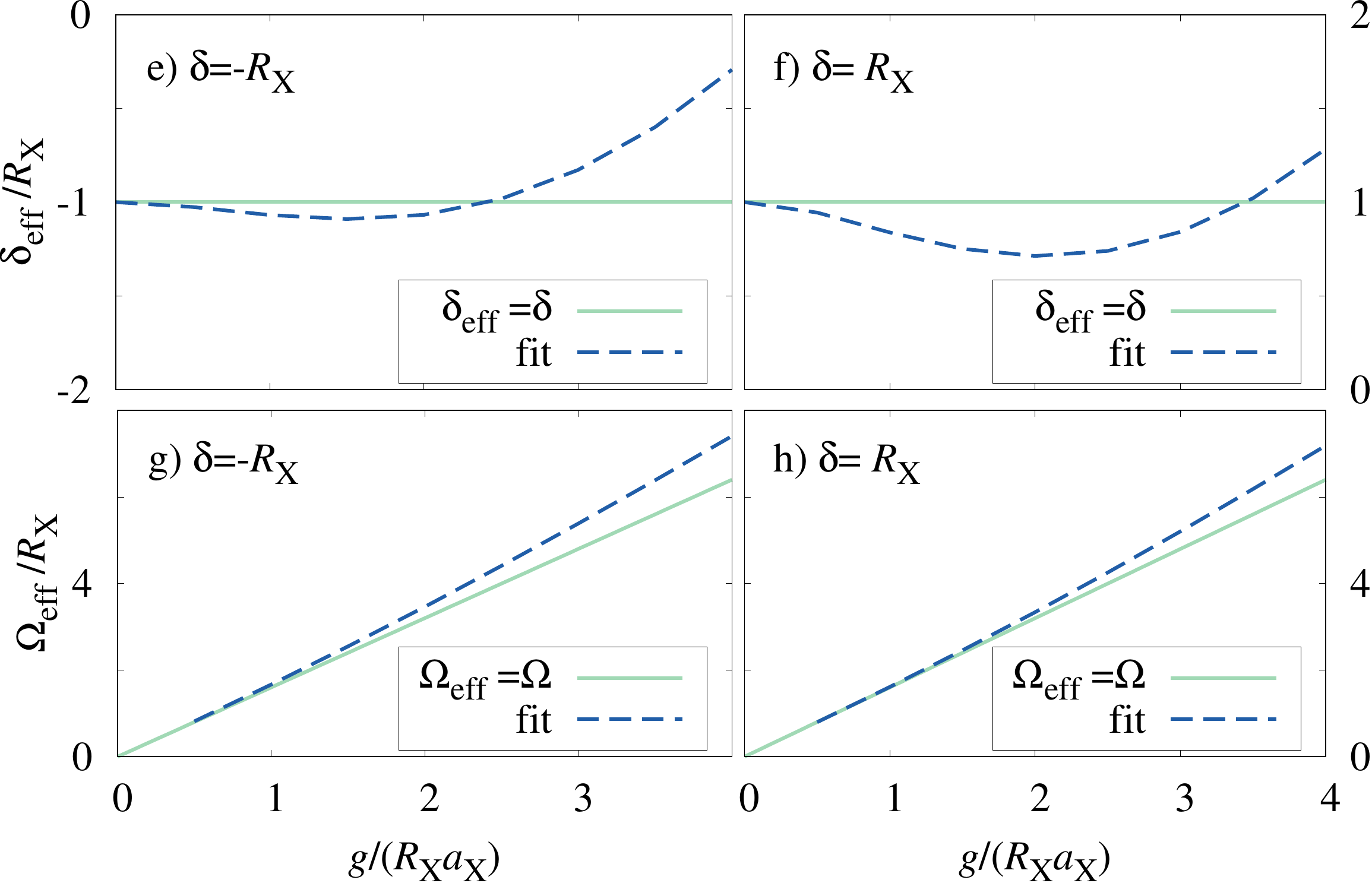}
\caption{Comparison between detuning $\delta$~\eqref{eq:detun} and Rabi splitting $\Omega$~\eqref{eq:rabic}, as defined in the renormalization procedure of Sec.~\ref{sec:renor}, and the respective  quantities $\delta_{\text{eff}}$ and  $\Omega_{\text{eff}}$ obtained by a least square fitting procedure described in App.~\ref{app:fitti}. Parameters are for a GaAs microcavity with a single quantum well ($d=0$), $m_2/m_1=0.25$, and $\ef=0$. }
\label{fig:impro}
\end{figure}
%
\section{Validity of the renormalization procedure beyond the \texorpdfstring{$g\ll\ax \Ryx$}{g<<ax Ex} limit}
\label{app:fitti}
We discuss here an improvement  of the renormalization procedure employed in the main text, to increase its accuracy beyond the weak coupling limit. 
In Sec.~\ref{sec:renor} we saw that in the weak coupling limit $g\ll\ax \Ryx$, defining the renormalized photon-exciton detuning $\delta$ as in Eq.~\eqref{eq:detun} and the Rabi splitting $\Omega$ as in Eq.~\eqref{eq:rabic}, enables one to recover the one-particle LP energy of the coupled oscillator model, Eq.~\eqref{eq:lpcou}.
Beyond weak coupling, the exciton wavefunction is strongly modified by light-matter coupling, thus impacting the detuning and the Rabi splitting. 
Here, we provide alternative definitions for the effective detuning $\delta_{\text{eff}}$ and Rabi splitting $\Omega_{\text{eff}}$ that coincide with the previous ones for $g\ll\ax \Ryx$, but whose validity
extends beyond this limit.
Comparing the two results  allows one to estimate the quantitative error made in our study of the evolution of the system phase diagram with increasing Rabi splitting $\Omega$, see Sec.~\ref{sec:compa}. 

To renormalize the theory, it is necessary to identify a  measurable quantity which can be used to define the renormalized quantities in the theory.
Ideally, the quantity we would use would be the photon energy.  However, this is not directly measurable, since the renormalization only occurs for a cavity which contains an active medium, and in that case, the photon mode is replaced by the strongly coupled polariton modes.  To circumvent this problem, as in Ref.~\onlinecite{Levinsen_arxiv2019}, we define the effective detuning $\delta_{\text{eff}}$ and Rabi splitting $\Omega_{\text{eff}}$ in a way analogous to an experimental procedure --- by fitting the polariton dispersion to a coupled oscillator model.

\begin{figure}
\centering
\includegraphics[width=1.0\columnwidth]{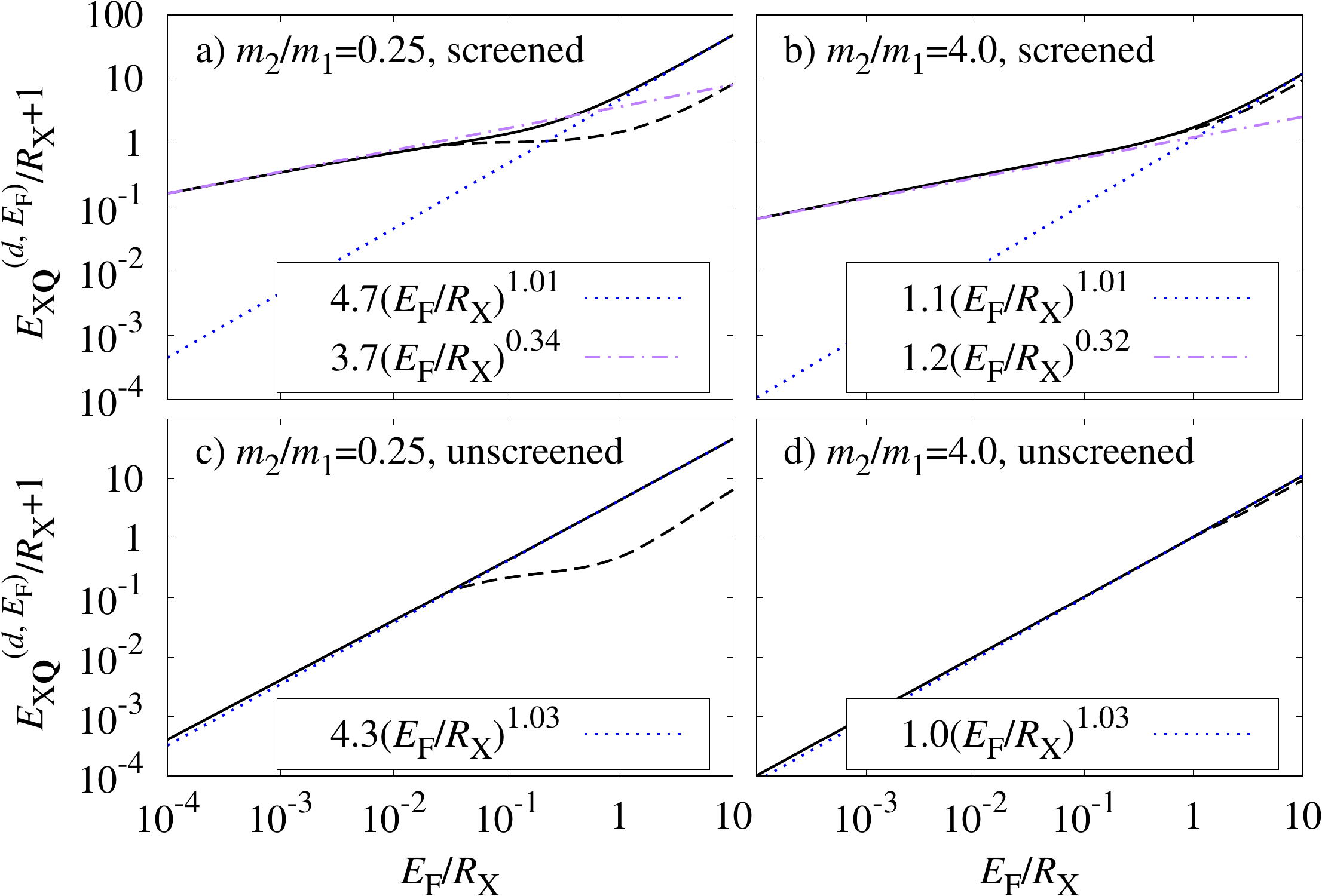}
\caption{Rescaled shift of  exciton energies $\exQdef - \exd$ at $\Q=\0$ (solid) and $\Q_{\text{min}}$ (dashed) as a function of density. Parameters are for a GaAs single quantum well ($d=0$, $\exdzero=-\Ryx$), two different mass ratios $m_2/m_1=0.25$ and $m_2/m_1=4$, and for both screened (top panels, $N_s=1$) and unscreened (bottom panels, $N_s=0$) interactions. Dot-dashed and dotted lines show the low- and high-density fittings, respectively.}
\label{fig:exenr}
\end{figure}

In particular, we employ a two-parameter least square fitting procedure to match the LP dispersion $E_{\Q}$ evaluated numerically from Eqs.~\eqref{eq:eigen} with the LP dispersion obtained by the coupled oscillator model~\eqref{eq:lpcou},
\begin{multline}
  \omega_{\text{LP}\Q}= \exd + E_g+\Frac{\delta_{\text{eff}} +
    \frac{\Q^2}{2m_{\text{C}}}+ \frac{\Q^2}{2(m_1+m_2)}}{2} \\
    -
  \frac{1}{2} \sqrt{\left(\delta_{\text{eff}} +
    \frac{\Q^2}{2m_{\text{C}}}- \frac{\Q^2}{2(m_1+m_2)}\right)^2 +
    \Omega_{\text{eff}}^2} \; ,
\label{eq:omelp}    
\end{multline}
where $\delta_{\text{eff}}$ and $\Omega_{\text{eff}}$ are fitting parameters.
In Fig.~\ref{fig:impro} we compare the results obtained for the fitting parameters $\delta_{\text{eff}}$ and $\Omega_{\text{eff}}$ with $\delta$ and $\Omega$ as defined in Eqs.~\eqref{eq:detun} and~\eqref{eq:rabic}, respectively.
In  panels a)-d) we fix the light-matter coupling $g$ and vary $\delta$, while in panels e)-h), we fix $\delta$ and vary $g$.
As expected, $\delta_{\text{eff}} \to \delta$ and $\Omega_{\text{eff}} \to \Omega$ when $g \ll \Ryx\ax$.
Moreover, we observe that the differences  $|\delta_{\text{eff}}-\delta|$ and $\Omega_{\text{eff}} - \Omega$ remain relatively small also when $g\gtrsim \Ryx\ax$.
These results allow us to estimate the size of the corrections that would arise from an improved renormalization scheme. We see that these appear small.  Nonetheless, there may be some changes in the results of Sec.~\ref{sec:compa}, when studying the phase diagram beyond the $g \ll \Ryx\ax$ regime.

\section{Exciton energies at finite \texorpdfstring{$\ef$}{EF}}
\label{app:phfra}
\begin{figure}
\centering
\includegraphics[width=0.9\columnwidth]{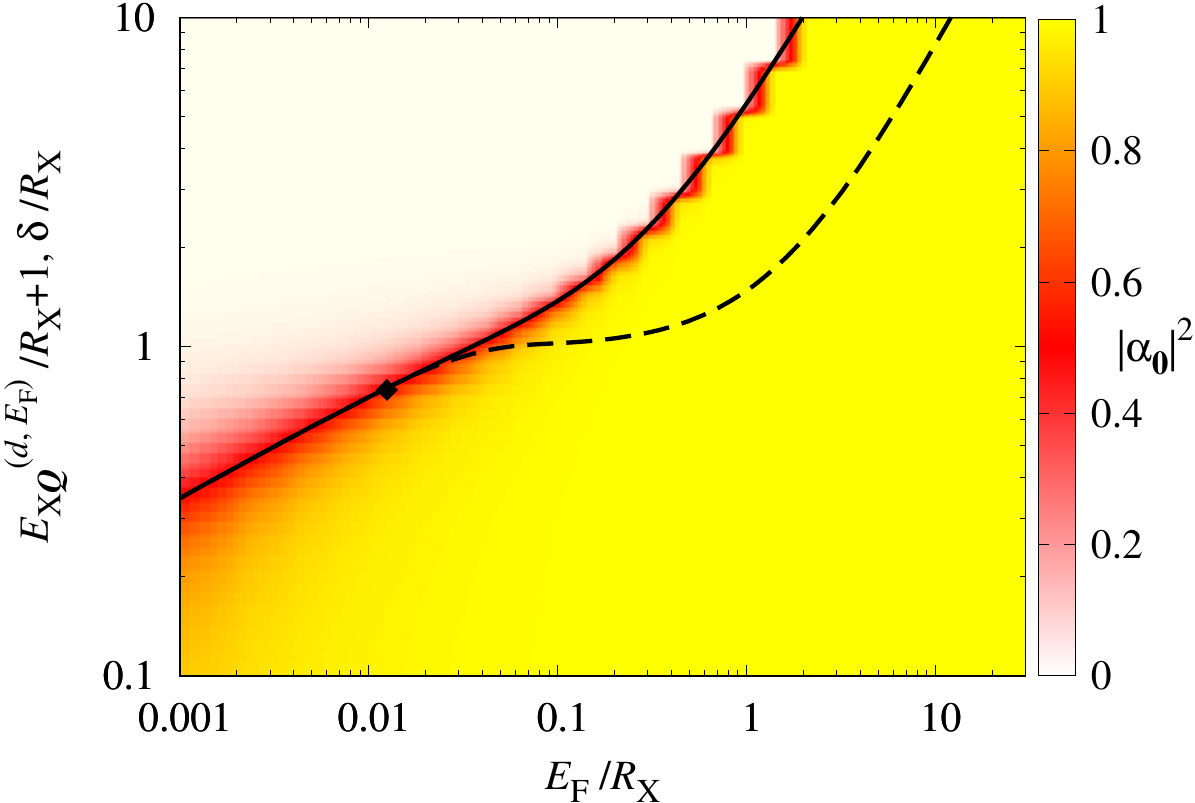}
\caption{Rescaled shift of  exciton energies $\exQdef - \exd$ at $\Q=\0$ (solid) and $\Q_{\text{min}}$ (dashed) as a function of the density. Parameters are for a GaAs single quantum well ($d=0$, $\exdzero=-\Ryx$), $m_2/m_1=0.25$, and screened interactions $N_s=1$. The diamond symbol represents  the Fermi energy $\efzero$ at which the two energies split as a consequence of a second-order transition where $\Q_{\text{min}}$ moves away from $\0$. The color map is the photon fraction $|\alpha_{\0}|^2$ of the  many-body polariton state at $\Q=\0$ for $\Omega=0.2\Ryx$. The color map is plotted against $E_F/\Ryx$ ($x$-axis) and detuning $\delta/\Ryx$ ($y$-axis).}
\label{fig:phfra}
\end{figure}
In Fig.~\ref{fig:exenr}, we compare the density dependence behaviour of the rescaled binding energies $\exQdef - \exd$ of the many-body exciton state at $\Q=\0$ (solid line) and at $\Q_{\text{min}}$ (dashed line) for different mass ratios $m_2/m_1=0.25,4$ and for both screened and unscreened interactions.
Note that while the dependence of $\exQzerodef$ on $\ef$ is sub-linear for small values of $\ef$ and screened interactions, it eventually becomes linear at large $\ef$.

In Fig.~\ref{fig:phfra} we plot $\exQdef - \exd$ as a function of density for a specific choice of parameters and  superimpose a color map of the photon fraction $|\alpha_{\0}|^2$ of the many-body $\Q=\0$ polariton state, as a function of $E_F$ and detuning $\delta$. The red region shows where the photon fraction is around $50\%$ indicating that the cavity photon energy is resonant with the many-body $\Q=\0$ exciton state --- see Eqs.~\eqref{eq:stron} and~\eqref{eq:delfi}.
As discussed is Sec.~\ref{sec:effec}, the photon energy shift at $\Q=\0$, $\wCQzeroef - \omega_{\text{C}\0}$, depends only weakly on $\ef$.
In particular, for the small value of $\Omega$ used in Fig.~\ref{fig:phfra} ($\Omega = 0.2\Ryx$), we expect that the $\ef$ dependence of the effective photon energy is negligible with respect to that of the exciton energy, $|\wCQzeroef - \omega_{\text{C}\0}| \ll |\exQzerodef - \exd|$. Thus, in this case, we expect that $\delta_{50\%} \simeq \exQzerodef - \exd$, which matches what  is observed in Fig.~\ref{fig:phfra}: The detuning $\delta$ at which resonance occurs (red region) coincides with the energy shift of the exciton, $\exQzerodef - \exd$ (solid line). 

\begin{figure}
\centering
\includegraphics[width=0.8\columnwidth]{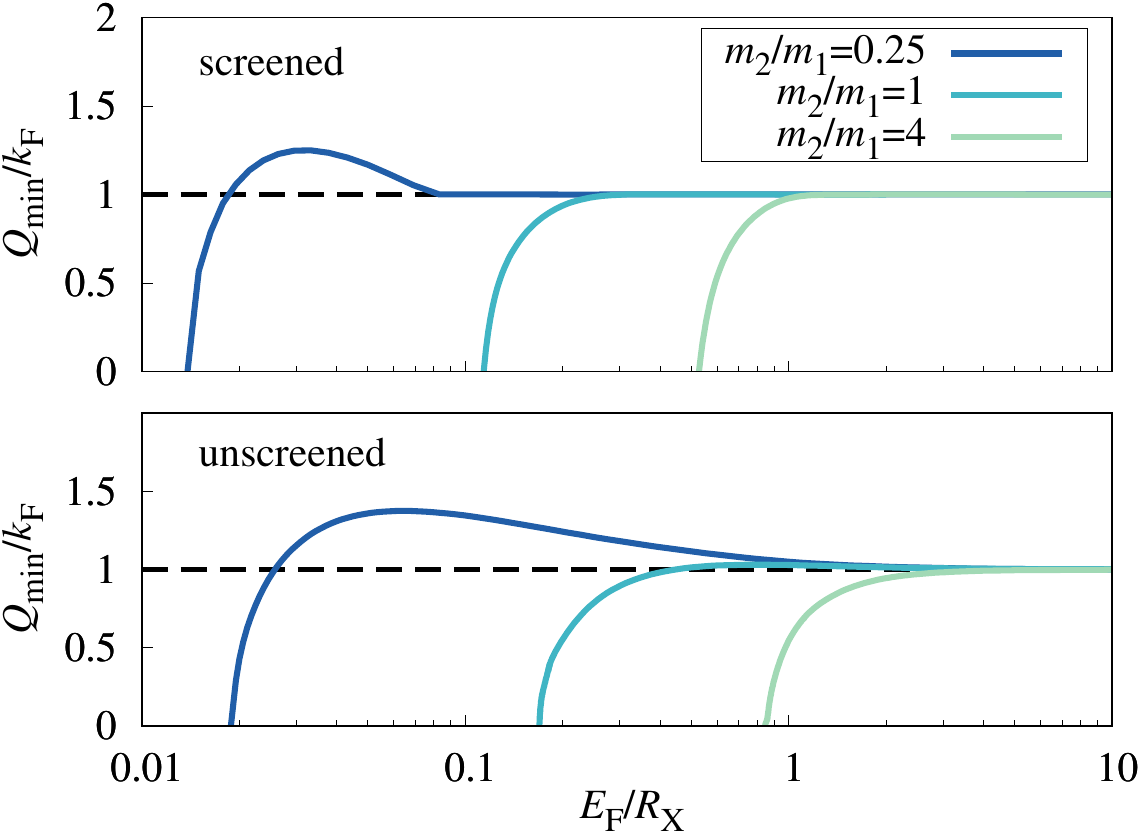}
\caption{Momentum $Q_{\text{min}}$ minimizing the many-body  exciton  energy $E_{\Q} = \exQdef$ obtained by solving 
Eq.~\eqref{eq:eige3}  as  a  function  of  the majority Fermi energy $\ef$ for  a  single  quantum  well  ($d=0$), and for different values of the mass  ratios $m_2/m_1$ as indicated. Top panel is screened, $N_s=1$, and bottom panel is unscreened, $N_s=0$.}
\label{fig:secon}
\end{figure}
\begin{figure}
\centering
\includegraphics[width=0.75\columnwidth]{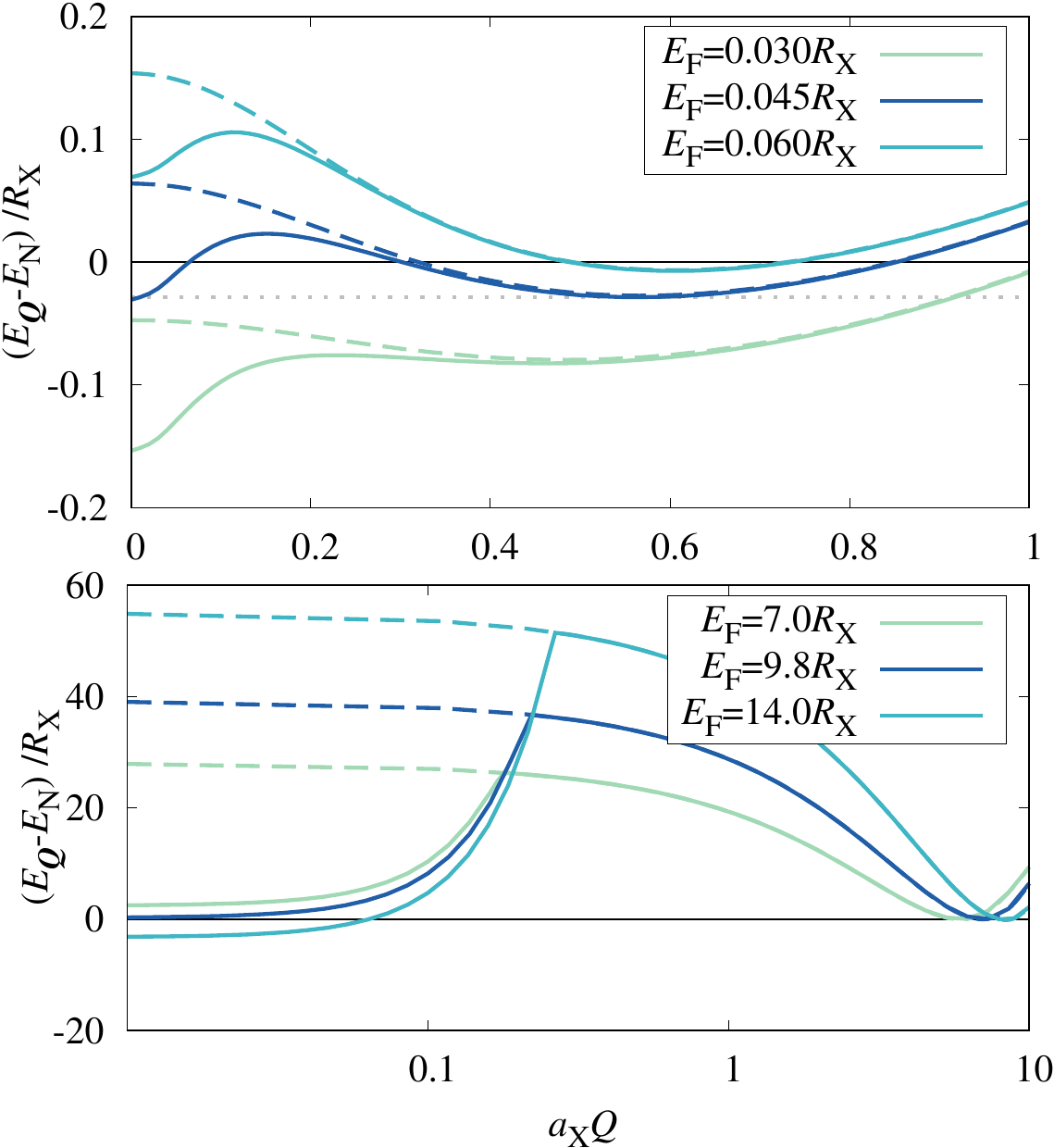}
\caption{Many-body polariton ground state energy $E_{\Q}$ with respect to the normal state energy $E_{\text{N}}$ (solid lines) versus momentum $Q$. Parameters are for a GaAs heterostructure with a single quantum well ($d=0$), mass ratio $m_2/m_1=0.25$, Rabi splitting $\Omega = 2\Ryx$, detuning $\delta=8\Ryx$ and screened interactions ($N_s=1$). Dashed colored lines are the many-body exciton energies $\exQdef$ evaluated in absence of the light-matter coupling, $\Omega=0$.The gray dotted line indicates where the minima at $Q=0$ and $Q\neq 0$ are equal. The top panel shows the 1$^{\text{st}}$ order SF-FF transition when increasing the system density, while the bottom panel shows the N-SF 1$^{\text{st}}$ order transition.}
\label{fig:enerq}
\end{figure}
\begin{figure}
\centering
\includegraphics[width=1\columnwidth]{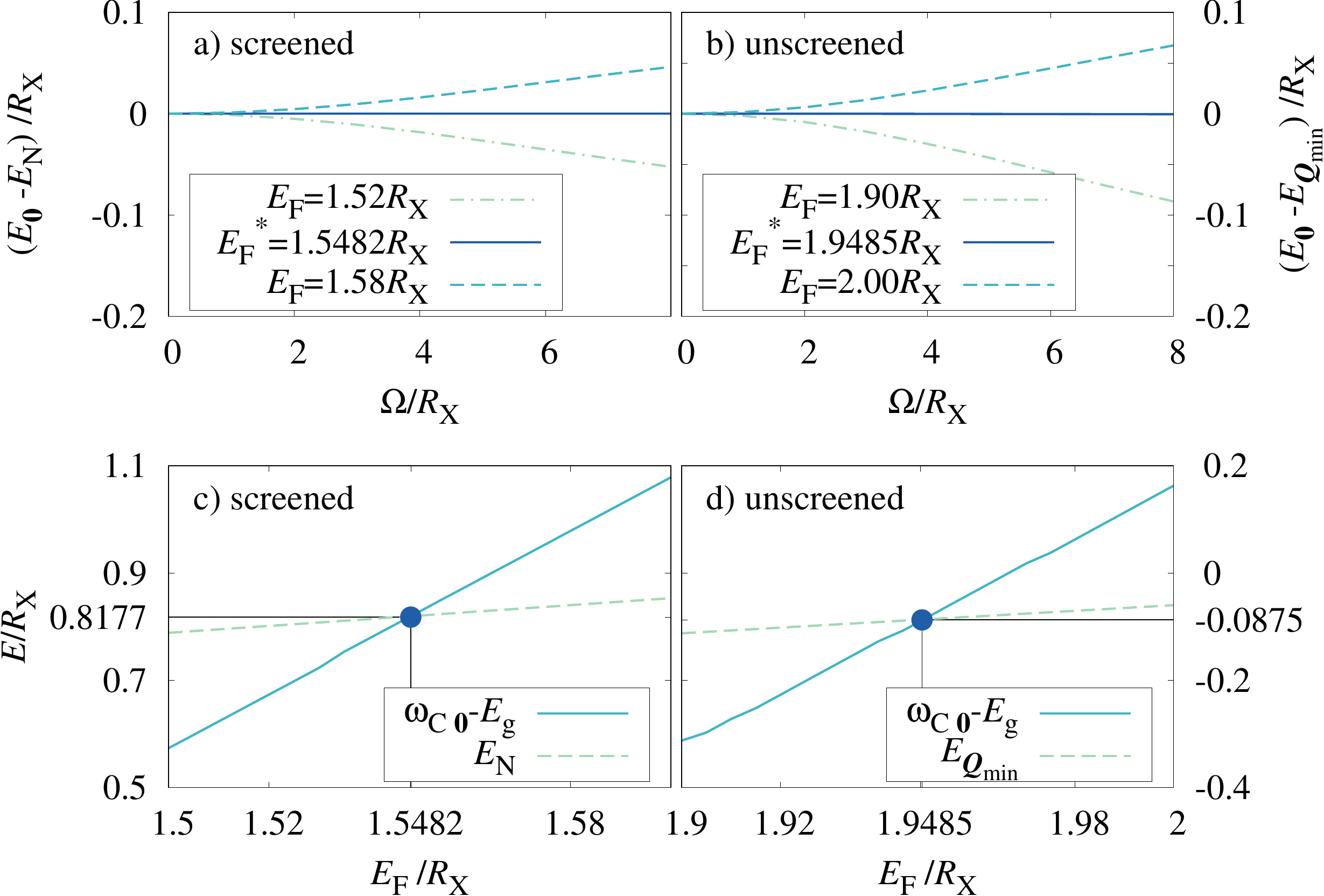}
\caption{Illustration of the dependence of energies on Rabi splitting close to 
$\ef^*$. The parameters are for a GaAs heterostructures with a single quantum well $d=0$, mass ratio $m_2/m_1=0.25$ and for screened (left panels) and unscreened (right panels) interactions. Top panels: energy difference between the many-body polariton SF energy $E_{\0}$ and the FF energy $E_{\Q_{\text{min}}}$ (top right) or between $E_{\0}$ and the normal state energy $E_{\text{N}}$ (top left). For each Fermi energy, $\ef$, the detuning is fixed according to Eq.~\eqref{eq:delts}, describing the SF-FF boundary at $\Omega=0$. Bottom panels: Photon energy $\omega_{\text{C}\0} - E_g$ satisfying Eq.~\eqref{eq:stco2} at $\Q=\0$ and $E_{\Q_{\text{min}}}$ --- for the values of $\ef$ considered in the plot and for screened interactions, $E_{\Q_{\text{min}}}$ coincides with the normal state energy $E_{\text{N}}$, i.e., $\Q_{\text{min}}=\kf \hat{\k}_{\text{F}}$.}
\label{fig:efsta}
\end{figure}
\section{\texorpdfstring{$1^{\text{st}}$ vs. $2^{\text{nd}}$}{1st vs 2nd} order transitions}
\label{app:order}
As shown  in Ref.~\onlinecite{Parish_EPL2011}, in the absence of the photon field the excitonic SF-FF transition is always second order. In Fig.~\ref{fig:secon} we show this by plotting the momentum $Q_{\text{min}}$ --- which minimizes the many-body exciton energy $E_{\Q}=\exQdef$ solution of Eq.~\eqref{eq:eige3} --- as a function of the Fermi energy of the majority species. We see that the transition from the SF $\Q=\0$ to the finite momentum FF phase is continuous. In addition, for screened interactions, when increasing the density further, $Q_{\text{min}}$ locks to precisely $\kf$ at the FF-N transition.

In the presence of a cavity field, the transitions SF-FF and SF-N are instead  first order. This is shown in Fig.~\ref{fig:enerq}, where we plot the energy of the polaritonic state vs $Q$.  
These data refer to the parameters of Figs.~\ref{fig:detun} and~\ref{fig:qminf}. In the top panel we show three curves  varying the majority species density close to the first SF-FF transition. We have taken a positive large value of the detuning at $\ef=0$, $\delta=8\Ryx$, such that the photon energy is far above the range of energies shown on this figure. Nevertheless, by comparing the many-body LP energy $\Q$-dispersion with that of the many-body exciton (dashed lines, corresponding to $\Omega=0$), we  observe significant effects of mixing between light and matter near $\Q=\0$ . The light-matter mixing at $Q_{\text{min}}$ is much smaller, around $|\alpha_{Q_{\text{min}}}|^2\simeq 10^{-6}$.  As a result, the energy shows two local minima which cross --- the signature of a first order transition.
For the N-SF transition that occurs at larger majority species density, bottom panel of Fig.~\ref{fig:enerq}, we observe that there is minimal coupling between matter and light both for the SF $\Q=0$ state (because the $\Q=\0$ exciton energy state is here pushed to very  high energies  by Pauli blocking) and for the normal state at $Q_{\text{min}}=\kf$ which has zero photon fraction.

\section{Origin of 
\texorpdfstring{$\ef^*$}{EF*} and \texorpdfstring{$\delta^*$}.}
\label{app:efsta}
We explain here the origin of the ``universal point'' $(\ef^*,\delta^*)$ found in the phase diagram of Fig.~\ref{fig:chang}. Remarkably, exactly at this point there is no $\Omega$ dependence of either the SF-FF transition (for unscreened interactions) or the SF-N transition (for screened interactions).
One way to understand the origin of this universal point is by comparing the many-body LP energy of the SF state at $\Q=\0$, $E_{\0}$, with that of the FF phase at $Q_{\text{min}}$, $E_{Q_{\text{min}}}$. The two energies clearly coincide at this $1^{\text{st}}$ order boundary (for screened interactions the FF phase may be replaced by the N phase if the density is large enough).
A limiting case of this boundary occurs when $\Omega \to 0$; in this limit the boundary occurs when 
\begin{equation}
  \delta = \exQdefmin - \exd\; ,
\label{eq:delts}  
\end{equation}
(assuming $\ef>\efzero$),
where $\exQdefmin$ is the many-body exciton (i.e., $\Omega=0$ case) energy of the FF phase for a majority species Fermi energy $\ef$.  This condition corresponds to a crossing between a photonic SF state and the excitonic FF state.  At non-zero $\Omega$, the SF state becomes polaritonic.

The existence of the special point $(\ef^*,\delta^*)$ corresponds to a point where this critical condition is not affected by light-matter coupling. To see this, we consider the following.
At each $\ef$, we can choose the detuning $\delta$ so as to satisfy Eq.~\eqref{eq:delts},  thus on the SF-FF boundary at $\Omega=0$. We then plot in the top panels of Fig.~\ref{fig:efsta}  $E_{\0}-E_{\Q_{\text{min}}}$, the energy difference between the LP energy at $\Q=\0$ and $\Q=\Q_{\text{min}}$, as a function of $\Omega$. We plot this energy difference
for different values of $\ef$. For $\ef<\ef^*$, this energy difference decreases with $\Omega$.  This means that  on increasing $\Omega$, the SF-FF boundary moves  to larger values of the detuning (see Fig.~\ref{fig:chang}). Conversely, if $\ef>\ef^*$, the energy difference increases with  $\Omega$, so the SF-FF boundary moves down to lower  detuning. Exactly at $\ef=\ef^*$,  we observe that $E_{\0}-E_{\Q_{\text{min}}}=0$, becomes exactly independent of $\Omega$.  As such, at this value of $\ef^*$, the critical detuning is $\delta^*$, independent of $\Omega$

Given the effective $\Omega$ independence seen at $\ef^*$, an alternative way of identify the value of $\ef^*$ and $\delta^*$ is by finding a condition for which the eigenenergy of the variational state becomes independent on the coupling to light.
To do this, following Ref.~\onlinecite{Levinsen_arxiv2019}, we rewrite the many-body eigenvalue problem of Eqs.~\eqref{eq:eigen} in terms of the renormalized photon energy $\omega_{\text{C}\Q} = \omega_{\text{C}\0} + \Q^2/2 m_{\text{C}}$~\eqref{eq:renor}, to give an expression which is independent of the UV cut-off. We thus separate out the divergent part of the relative wave-function $\varphi_{\k\Q}$,
\begin{equation}
  \varphi_{\k\Q} = \beta_{\k\Q} + \Frac{g \alpha_{\ve{Q}}}{E_{\Q}  - \xi_{\k\Q}}  \; ,
\label{eq:betad}
\end{equation}
and rewrite~\eqref{eq:eigen} in the following equivalent forms:
\begin{widetext}
\begin{subequations}
\label{eq:eqeig}
\begin{align}
\label{eq:eig1p}
  &\left(E_{\Q}  - \xi_{\k\Q}\right) \beta_{\ve{k}\ve{Q}} = -
  \Frac{1}{\area}\sum_{\ve{k'}>\kf} V_{\ve{k}-\ve{k'}}  \beta_{\ve{k'}\ve{Q}} +  \Frac{g \alpha_{\ve{Q}}}{\area}\sum_{\ve{k'}>\kf} \Frac{V_{\ve{k}-\ve{k'}}}{-E_{\Q}  + \xi_{\k'\Q}}\\
    &\left[E_{\Q}  - \omega_{\text{C}\vect{Q}} + E_g + \Frac{g^2}{\area} \left( \sum_{\ve{k}>\kf} \Frac{1}{-E_{\Q}  + \xi_{\k\Q}} - \sum_{\ve{k}} \Frac{1}{-\exd + \epsilon_{\k,1}+ \epsilon_{\k,2}}\right)\right]
    \alpha_{\ve{Q}} = \Frac{g}{\area} \sum_{\ve{k}>\kf}
    \beta_{\ve{k}\ve{Q}} \; .
\label{eq:eig2p}
\end{align}
\end{subequations}
\end{widetext}
All sums are now convergent.  For the solution of these equations to be independent of light-matter coupling means the $E_{\Q}$ must match the solution at $g=0$, i.e.,
\begin{equation}
    E_{\Q} = \omega_{\text{C}\vect{Q}}-E_g \;.
\label{eq:decou}
\end{equation}
This condition corresponds to the system energy $E_{\Q} + E_g$ coinciding with $\omega_{\text{C}\Q}$, the energy of the photon mode at $\ef=0$.
Using Eq.~\eqref{eq:decou} in Eq.~\eqref{eq:eig2p}, we obtain the following equation to define $\ef^*$:
\begin{multline}
  \sum_{\ve{k}>\kf} \Frac{1}{-E_{\Q}  + \xi_{\k\Q}} - \sum_{\ve{k}} \Frac{1}{-\exd + \epsilon_{\k,1}+ \epsilon_{\k,2}}\\
     = \Frac{1}{g \alpha_{\ve{Q}}} \sum_{\ve{k}>\kf}
    \beta_{\ve{k}\ve{Q}}\; .
\label{eq:stcon}    
\end{multline}
Note that this condition is indeed independent of  $g$.  To see this, we formally invert Eq.~\eqref{eq:eig1p}, to give  $\beta_{\ve{k}\ve{Q}}$:
\begin{equation}
  \beta_{\ve{k}\ve{Q}} = g \alpha_{\ve{Q}} \sum_{\ve{k'}>\kf} ({\mathbb{M}_{\Q}}^{-1})_{\k,\k'} L_{\k'\Q}\; ,
\end{equation}
where the matrix $\mathbb{M}_{\Q}$ and vector $L_{\Q}$ in relative momentum space are defined respectively as
\begin{align}
  (\mathbb{M}_{\Q})_{\k,\k'} &= \left(E_{\Q}  - \xi_{\k\Q}\right) \delta_{\k,\k'} + V_{\ve{k}-\ve{k'}} \\
  L_{\k\Q} &= \sum_{\ve{k'}>\kf} \Frac{V_{\ve{k}-\ve{k'}}}{-E_{\Q}  + \xi_{\k'\Q}} \; .
\end{align}
We thus find that Eq.~\eqref{eq:stcon} is independent of both $g$ and $\alpha_{\ve{Q}}$:
\begin{multline}
  \sum_{\ve{k}>\kf} \Frac{1}{-E_{\Q}  + \xi_{\k\Q}} - \sum_{\ve{k}} \Frac{1}{-\exd + \epsilon_{\k,1}+ \epsilon_{\k,2}}\\
     = \sum_{\ve{k}>\kf}
    \sum_{\ve{k'}>\kf} ({\mathbb{M}_{\Q}}^{-1})_{\k,\k'} L_{\k'\Q} \; .
\label{eq:stco2}    
\end{multline}

In addition to satisfying Eq.~\eqref{eq:stco2},
$\ef^*$ lies on the SF-FF (SF-N) boundaries for unscreened (screened) interactions and, thus, it also has to lie on the boundary at $\Omega=0$. With this in mind, we plot in the bottom panels of  Fig.~\ref{fig:efsta} the energy $\omega_{\text{C}\0}- E_g = E_{\0}$ obtained by solving Eq.~\eqref{eq:stco2} at $\Q=\0$ as a function of 
$\ef$. From the crossing of this curve with that of the FF state in the absence of light, i.e., the FF exciton energy  $\exQdefmin$ (or, for the screened case, the normal state energy $E_{\text{N}}$), we recover the value of $\ef^*$.
The corresponding value of the detuning $\delta^*$ is given by Eq.~\eqref{eq:delts} for $\ef=\ef^*$, i.e., $\delta^* =E_{\text{X}\Q_{\text{min}}}^{(d,\ef^*)} - \exd$.
We thus find $(\ef^*,\delta^*)\simeq(1.55\Ryx,1.82\Ryx)$ (for screened interactions) and  $(\ef^*,\delta^*)\simeq(1.95\Ryx,0.91\Ryx)$ (unscreened interactions). 

Finally, we remark that the $g$ independence at $\ef^*$ does not imply that light and matter are fully decoupled at this point. Indeed, the photon frequency depends on the active medium through the process of renormalization. However, precisely at $\ef^*$, the photon self energy arising due to the light-matter interaction only contains the term that appears in Eq.~\eqref{eq:renor}, while all other terms cancel.
Given the general arguments that led us to determining the point $(\ef^*,\delta^*)$, it is likely that it persists as a special point in the photon self energy also beyond the variational approach used in this work.


%

\end{document}